\documentclass[prd,preprintnumbers,reprint,twocolumn,notitlepage]{revtex4-1} 
\usepackage{dcolumn,bm}
\usepackage{amsmath,amssymb,amsfonts,graphics,graphicx,amscd,amsfonts,color}
\usepackage[colorlinks=true,allcolors=blue]{hyperref}
\usepackage[all]{hypcap}
\usepackage{verbatim}
\DeclareMathAlphabet{\mathpzc}{OT1}{pzc}{m}{it}
\definecolor{Violet}{rgb}{0.5,0,1}
\definecolor{darkgreen}{rgb}{0,0.64,0}
\newcommand{\CC}{$\Lambda\,$}
\newcommand{\RI}{Region I}
\newcommand{\RII}{Region II}
\newcommand{\RIII}{Region III}
\newcommand{\RIV}{Region IV}
\newcommand{\RV}{Region V}
\newcommand{\bwt}{\begin{widetext}}
\newcommand{\ewt}{\end{widetext}}
\newcommand{\dOmt}{\,d\Omega^{2}_{\it 2}}
\newcommand{\dOmd}{\,d\Omega^{2}_{\it d}}

\newcommand{\ein}{\emph{In} }
\newcommand{\eout}{\emph{Out} }

\allowdisplaybreaks[4]
\date{\today}
\begin{document}
\title{de Sitter Harmonies: Cosmological Spacetimes as Resonances}
\author{Jonathan Maltz$^{a,b}$} 
\affiliation{${}^a$ Center for Theoretical Physics and Department of Physics, University of California at Berkeley, Berkeley, California, 94720,USA}
\affiliation{${}^b$ Stanford Institute for Theoretical Physics, Stanford University, Stanford, California 94305, USA}
\preprint{SU-ITP-16/16}
\begin{abstract}
The aim of this work is to provide the details of a calculation summarized in the recent paper by Maltz and Susskind which conjectured a potentially rigorous framework where the status of de Sitter space is the same as that of a  resonance in a scattering process. The conjecture is that transition amplitudes between certain states with asymptotically supersymmetric flat vacua contain resonant poles characteristic metastable intermediate states. A calculation employing constrained instantons is presented that illustrates this idea.
\end{abstract}
\maketitle
\tableofcontents
\section{Introduction and Motivations}
\hspace{0.1in}String/M-theory is the leading candidate for a formalism of quantum gravity \cite{Polyakov:1981rd,Green:1987sp,Green:1987mn,Polchinski:1998rq,Polchinski:1998rr,Schwarz1999107,klebanov:28,Aharony:1999ti,Tong:2009np}, having had many successes in providing an ultraviolet completion of gravitational phenomena that are described to high experimental precision in the infared  by general relativity (GR) \cite{Einstein:1915by,Einstein:1915ca}. Reproducing the spectrum of ten-dimensional supergravity at low energies, providing controlled calculations of black hole microstate counting \cite{Strominger199699}, and introducing new notions into physics such as holographic complementarity, Matrix model descriptions of gravity \cite{Banks:1996vh,Aharony:2008ug,Ishibashi:1996xs}, and the AdS/CFT correspondence (gauge/gravity duality)\cite{Maldacena:1997re}. In describing cosmological spacetimes however, the theory is in a deep morass and descriptions reduce to Jabberwocky. 


Starting with supernova Ia measurements in 1987 \cite{Riess:1998cb,Perlmutter:1998np,Spergel:2003cb} and concurrent cosmic microwave background measurements \cite{Smoot:1992td,Perlmutter:1998np,Fixsen:1996nj} it has become apparent that the Universe's expansion is accelerating. Our explanation for this within the \textbf{\CC}-CDM model of cosmology is that the mass-energy density of the Universe is dominated by dark energy in the form of a small cosmological constant (\CC) \cite{Steinhardt:2006bf,Ade:2015xua,Peebles:2002gy,Hinshaw:2006ia,0067-0049-208-2-20} \footnote{More recent experimental data have only strengthened this result with WMAP's final combined best fit (WMAP + eCMB + BAO + H0) \cite{Hinshaw:2006ia,0067-0049-208-2-20} and Planck satellite data \cite{Ade:2015xua} putting the Hubble constant at $69.32\pm 0.80\frac{km}{s\cdot Mpc}$ and $67.80\pm 0.77\frac{km}{s\cdot Mpc}$ respectfully. This Hubble constant corresponds to a \CC of $10^{-47} GeV^{4}$ \cite{Tegmark:2003ud} or $\sim 10^{-122}$ in Planck units \cite{Barrow:2011zp}; yielding a dark energy percentage of $(69.2 \pm 1.2)\%$ from \textbf{Planck} 2015 data \cite{Ade:2015xua}.}.  This second exponential expansion phase, separate from the initial inflationary epoch \cite{Peebles:2002gy,Baumann:2009ds} that occurred just after the big bang \cite{Riess:1998cb,Perlmutter:1998np,Dodelson:1282338,weinberg2008cosmology}, implies that our Universe is best described as being  asymptotically de Sitter (dS) \cite{Nagamine:2002wi,Busha:2003sz,Dodelson:1282338,Frieman:2008sn,weinberg2008cosmology} from $10^{-33}$s after the big bang until far into the future. If string theory is going to directly address the issues of cosmology it is necessary to formulate a quantum definition of asymptotically dS spacetimes within string theory.

Computation of observable quantities in string theory typically relies on computing asymptotic states on what has been colloquially referred to \emph{asymptotically cold} backgrounds \cite{Susskind:2007pv} such as symptotically anti de Sitter (AdS) or asymptotically flat spacetimes i.e. the energy density and therefore fluctuations of the geometry go to 0 asymptotically or at the boundary where applicable, and gravity decouples. Because of the exponential expansion of the spacetime, dS possesses cosmological horizons. This implies that only a portion of the spacetime is ever accessible to any given observer and there is no asymptotically cold boundary region on which to define correlation functions \cite{Dyson:2002nt}. The region within the observer's horizon, referred to as the observer's causal patch \cite{Dyson:2002pf,Dyson:2002nt}, possesses a finite entropy and temperature \cite{Gibbons:1977mu,Hawking:2000da}. The finite entropy of the causal patch suggests that the causal patch of dS does not support exact states on its own and should be described by a large finite discreet  spectrum of states, which is incompatible with a continuum CFT description \cite{Dyson:2002nt} and the dS  symmetries \cite{Susskind:2003kw,Dyson:2002nt}\footnote{ There has been a great deal written in the literature on how to deal with dS spacetimes within the realm of string theory, far too much to fully cite but an incomplete list of the relevant works includes \cite{Hawking:2000da,Witten:2001kn,Strominger:2001pn,Strominger:2001gp,Bousso:2001mw,Maloney:2002rr,Goheer:2002vf,Kachru:2003sx,Freivogel:2004rd,Alishahiha:2004md,Parikh:2004wh,Banks:2006rx,Freivogel:2005vv,Freivogel:2006xu,Susskind:2007pv,Sekino:2009kv,Freivogel:2009rf,Dong:2010pm,Anninos:2010gh,Harlow:2010my,Anninos:2011ui,Harlow:2011ke,Anninos:2011jp,Harlow:2011az,Anninos:2012ft,Harlow:2012,Anninos:2012qw,Maltz:2012zs,Ng:2012xp,Maltz:2013}.}. 

Finally, string/M-Theory possesses a vast set of vacuum solutions known as the string theory landscape, with estimates of $\sim 10^{500}$ vacua \cite{Bousso:2000xa,Susskind:2003kw,Douglas:2003um,1982Natur.295..304G,1983veu..conf..251S,Linde1986395,doi:10.1142/S0217732386000129,Bousso:2007gp}. The most well-understood subset of these solutions is referred to as \emph{The Moduli Space of Supersymmetric Flat Vacua} (Supermoduli space) which are continuously connected to the five perturbative string theories \cite{Susskind:2003kw,Brown:1988kg,Susskind:2003kw,Bousso:2000xa}. Vacua in supermoduli space are supersymmetric preserving compactifications with $V(\varphi_n)=0$, ($\Lambda =0$). At low enough energies these moduli can be approximated by massless scalar fields that are under the influence of an effective potential $V(\varphi_n)$. Vacua are local minima of $V(\varphi_n)$ with $\Lambda$ equal to the value of the minima. Moving through moduli space means varying the dynamical moduli of the compactification, which changes the value of the effective fields $\varphi_n$ \footnote{Once we move off the 0 of the potential we must include contributions of energy from the  four form fluxes $F_{(4)}$ to give a non-zero vaccum energy [$F^{2}_{(4)}$]. The number and types of D-branes charged under these $F_{(4)}$ along with the various compactifications on which the fluxes are wrapped gives discreetum of vacua yielding the $10^{500}$ solutions. The $F_{(4)}$s and other fluxes also contribute to the effective potential and parametrize points on the landscape \cite{Brown:1988kg,Susskind:2003kw,Bousso:2000xa}}. Minima of the potential where $V(\varphi)\neq 0$ are obtained nonperturbatively.    dS vaccua, those with positive \CC, are in the landscape \cite{Kachru:2003aw,Kachru:2003sx}; however they are unstable to  vacuum decay  via Coleman de Luccia (\textbf{CDL}) tunneling \cite{Coleman:1977py,Callan:1977pt,Coleman:1980aw,Freivogel:2004rd,Freivogel:2006xu, Douglas:2003um} to flat or AdS vacua \footnote{Vacua in supermoduli space are marginally stable to vacuum decay \cite{Susskind:2003kw} and the AdS vacua crunch.}. The CDL decay complicates the structure of timelike future infinity $\mathcal{I}^{+}$ of dS, changing it to a history dependent fractal structure of many different types of bubbles of different cosmological constants \cite{Sekino:2010vc,Bousso:2011up}  
\footnote{The CDL decay of the dS manifests itself as bubbles of lower \CC forming in the dS \cite{Goheer:2002vf}, as the fields locally tunnel from one vacua to a lower one. A homogeneous tunneling of the entire dS to a true vacuum is prevented since in this case the fields acquire an infinite mass and act as classical coordinates, preventing tunneling \cite{Freivogel:2004rd}. Depending on the decay rate and Hubble constant of the ancestor dS, the coordinate volume of $\mathcal{I}^{+}$ will eventually be dominated by a fractal of bubbles possessing lower \CC s, but the proper volume will still be dominated by the ancestor dS; i.e., the ancestor dS inflates faster than the it is being eaten up by true vacuum allowing this process to proceed to infinitum. This leads to the eternal inflation scenario \cite{Linde1986395,doi:10.1142/S0217732386000129,Guth:2007ng,Freivogel:2006xu}. } in a quantum superposition.

The decay to \emph{Hats}, Friedmann-Roberston-Walker (FRW) bubbles of vacua in the supermoduli space, \CC$=0$, provides an opportunity to define a rigorous framework for dS. This conjectured framework known as FRW/CFT \cite{deBoer:2003vf,Freivogel:2004rd,Freivogel:2006xu,Susskind:2007pv,Sekino:2009kv,Harlow:2010my,Harlow:2011ke,Harlow:2012,Maltz:2012zs,Maltz:2013}\footnote{The conjecture being that in a parent dS that contains a \emph{Hat} on $\mathcal{I}^{+}$, the dual CFT of the \emph{hat} not only contains information of the FRW bulk but also of the portion of the parent dS in the FRW bubble's past light cone. Such a FRW bubble is marginally stable against CDL decay and possesses the proper asymptotic and entropy properties to describe exact CFT correlators. Invoking horizon complimentary \cite{Gibbons:1977mu,Susskind:1993if,Bigatti:1999dp,Banks:2001yp,Bousso:2002ju} across the dS horizon at the boundary of this light cone, the Hawking radiation coming off this horizon contains the information (albeit highly scrambled) of  anything that passed through it (the rest of the  multiverse)\cite{Gibbons:1977mu,Freivogel:2004rd} and hence the interior of the bubble contains this information as well. Cosmological horizons are of the type found in Rindler space \cite{Gary:2013oja} and would not suffer from Firewall issues resulting from the AMPS paradox \cite{Almheiri:2012rt,Braunstein:2009my}. In the  FRW/CFT framework the region analogous to the AdS/CFT boundary is the late time sky of the \emph{Hat} (topologically an $\mathbb{S}^{2}$) for a four dimensional hat, labeled $\Sigma$ in Fig. \ref{CDL}, which is spacelike infinity of the FRW bubble. The $O(3,1)$ symmetry of the \emph{hat}'s spacelike (EAdS$_{3}$) slices acts as 2D conformal transformations on $\Sigma$ \cite{Susskind:2007pv}.  It is conjectured in \cite{Freivogel:2006xu} that there is a nonunitary euclidean CFT on $\Sigma$ composed of a matter CFT with large positive central charge that depends on the ancestor dS's \CC and a two-dimensional gravitational sector described by a timelike Liouville field of compensating negative central charge along with ghost fields. The FRW/CFT framework is a holographic Wheeler-de Witt theory \cite{Freivogel:2004rd,Freivogel:2006xu,Susskind:2007pv,Sekino:2009kv,Harlow:2010my} and can be viewed as a dimensionally reduced dS/CFT, which is UV complete in the hat \cite{Susskind:2007pv,Sekino:2009kv,Harlow:2010my}.}.
In this work we provide the technical details of the computation inspired by FRW/CFT and summarized in \cite{Maltz:2016iaw}  to define a transition amplitude between supersymmetric flat vacua and show that resonant poles that we associate with dS metastable states exist in its spectral representation. 
To show this we consider a configuration looking like a time-symmetric slice of the dS vacuum and evolve the state in a time symmetric manner to yield the past and future infinity boundaries, which as previously stated are fractal superpositions containing an infinite number of hats as well as other vacua. Picking a past and future hat and invoking the gauge choice that they nucleate at the spatial center of a causal diamond, we define a transition amplitude and compute a spectral representation for this transition. This requires constructing a deformation of the CDL spacetime, which we will refer to as a constrained CDL instanton.  This spacetime, which is constructed via the Barrab\`es-Israel null junctions conditions \cite{PhysRevD.43.1129}, has the status of a constrained instanton \cite{Frishman:1978xs,Affleck:1980mp,Nielsen:1999vq} and is the result of the CDL instanton equations with a constraint that the FRW regions are separated in the dS region by a fixed proper time; see Fig \ref{constrainedCDL}. A regulated action is computed for this spacetime and a path integral for the transition amplitude  is performed using the  minisuperspace approximation in the thin-wall limit. Here the path integral over all deformations of the metric is constrained to only varying the time between the bubbles. Fourier transforming this amplitude with respect to this time  in order to get a spectral representation,  we find that the spectral representation contains resonant poles. We associate these poles to dS intermediate states. The idea that dS might be viewed as resonance has been suggested before in \cite{Susskind:2003kw,Freivogel:2004rd,Freivogel:2006xu}; however there is to the author's knowledge no explicit calculation to establish dS as a resonance or direct computation of the pole in the literature. We will present the details of one in this paper.


This paper is organized as follows: first in Sec. \ref{CDLsec} we introduce dS and  CDL instanton spacetime  \cite{Freivogel:2004rd,Freivogel:2006xu,Guth:2007ng}. In Sec. \ref{transamp} we define the transition amplitude and spectral representation \footnote{Reviews of spectral representations can be found \cite{goldberger2004collision,0521670535,0201503972,2008AIPC.1077...31R}.}. In Sec. \ref{calcul} we motivate the calculation and action prescription. Sections \ref{onepone}-\ref{totsactionandpole} contain the main bulk of the paper where we first compute the amplitude in $1+1$ Liouville gravity and $3+1$ Einsteinian gravity in order to establish the existence of the pole. $1+1$ dimensions, the Gauss-Bonnet theorem implies  that the boundary contributions may be neglected and the regulation of the action is simplified. In Sec. \ref{threepone} we compute the action in $3+1$ Einsteinian gravity taking into account the boundary terms. In the discussion we interpret this result and discuss its implications as well as present our conclusions. In Appendix \ref{geom} an explicit construction of the \emph{constrained} CDL spacetime employing the null junction conditions is presented. In \ref{stichjust},  an argument justifying the proposed integration region is presented.  Finally in \ref{geodchris} we give some useful relations for the geometry.
  
\section{de Sitter Space and the Coleman de Luccia Amplitude}\label{CDLsec}

\hspace{0.1in}de Sitter space is a maximally symmetric solution of the Einstein field equations \cite{weinberg:1972,Misner:1974qy,0226870332,carroll2003spacetime,2006eins.book..120M,Anninos:2012qw},
\begin{equation}\label{einsteineqn}
G_{\mu\nu}=R_{\mu\nu}+\frac{1}{2}R g_{\mu\nu} + \Lambda g_{\mu\nu} =0,
\end{equation}
where the cosmological constant is given by $\Lambda$ yielding a dS radius of $l_{dS}=\sqrt{\frac{3}{\Lambda}}$; for our Universe $\Lambda \cong 1.7 \times 10^{-121} \sim 1/t_U \sim 10^{-122}$ in Planck units \cite{Barrow:2011zp,Ade:2015xua}\footnote{We  employ Planck units ($G_N ,\hbar,c,k_{b},\frac{1}{4\pi\epsilon_{0}} =1$) in this paper.}.

\begin{figure}[ht]
\begin{center}
\includegraphics[scale=0.5]{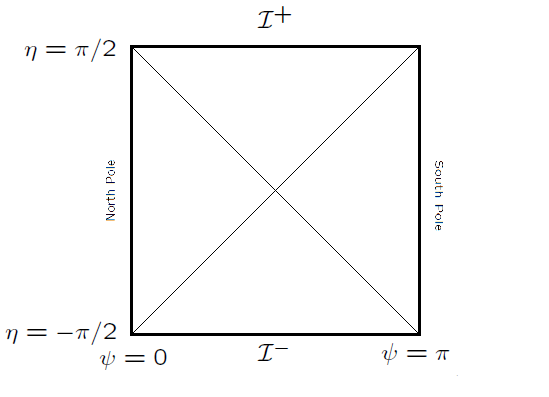}
\caption{Penrose diagram of de Sitter space. The north and south poles of the $\mathbb{S}^3$ are at $\psi = 0$ and $\psi =\pi$ respectfully. Timelike future infinity $\mathcal{I}^{+}$ is the line at conformal time $\eta=\pi/2$ and similarly timelike past infinity $\mathcal{I}^{-}$ is located at $\eta=-\pi/2$. The diagonal lines represent the horizons of the static patch. Note that timelike observers can only access a portion of the space irrespective of their starting point.}\label{desitter}
\end{center}
\end{figure}

Asymptotically dS spacetimes (cosmological spacetimes) add to (\ref{einsteineqn}) a stress tensor to describe the matter and radiation content of the Universe
\begin{equation}\label{einsteineqngen}
G_{\mu\nu}=R_{\mu\nu}+\frac{1}{2}R g_{\mu\nu} + \Lambda g_{\mu\nu} = \kappa T_{\mu\nu}.
\end{equation}
Solving (\ref{einsteineqn}), the metric for dS, written in global coordinates \footnote{When expressed in flat slicing coordinates, $ds^{2} = -dt^2 +e^{2\sqrt{\frac{\Lambda}{3}}t}\big(d\rho^2 +\rho^{2}\dOmt\big) $, the exponential expansion of the space becomes apparent. Here $d\,\Omega^{2}_{\it 2} = d\theta^{2} +\sin^{2}{\theta}\,d\phi^{2}$ is the metric for the unit $\mathbb{S}^{2}$. } is
\begin{equation}\label{globaldsmet}
ds^2=-dt^2+\frac{3}{\Lambda}\cosh^{2}{\Bigg[\sqrt{\frac{\Lambda}{3}}t\Bigg]}\big(d\psi^{2}+\sin^{2}{\psi}\dOmt\big).
\end{equation}
Using the relation
\begin{equation}\label{globaltimetoconftime}
\tanh{\Bigg[\sqrt{\frac{\Lambda}{3}}\frac{t}{2}\Bigg]}=\tan{\Big[\frac{\eta}{2}\Big]},
\end{equation}
 \footnote{The relation $\cosh{\Big[\sqrt{\frac{\Lambda}{3}}t\Big]}=\frac{1}{\cos{\eta}}$ also produces the same coordinate change.}, we can reexpress (\ref{globaldsmet}) in conformal time coordinates 
\begin{equation}\label{dsMetconf1}
ds^2_{dS} = \frac{3}{\Lambda \cos^{2}{\eta}}\Big\{-d\eta^2 + d\psi^2 + \sin^{2}{\psi}\dOmt\Big\},
\end{equation}
with $-\pi/2\leq\eta\leq\pi/2$ and $0\leq\psi\leq\pi$. These are the coordinates generally used to label the Penrose diagram for dS, shown in Fig. \ref{desitter}.
Pure dS (\ref{globaldsmet}) can be regarded as a 4d hyperboloid, $-(X^{0})^{2} + \sum^{4}_{i=1}(X^{i})^{2} = l^{2}_{dS} $, embedded in 5d Minkowski space $ds^2 = -(d\,X^{0})^2 +\sum^{4}_{i=1}(d\,X^{i})^2$ \cite{carroll2003spacetime,2006eins.book..120M,Anninos:2012qw}.

\begin{figure}[ht]
\begin{center}
\includegraphics{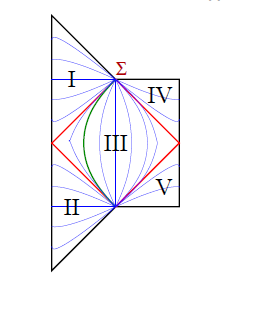}
\caption{The Penrose diagram of the Lorentzian continuation of CDL instanton solution \cite{PhysRevD.21.3305,Freivogel:2004rd,Freivogel:2006xu}. Regions \RI\, and \RII\, are open (k = -1) FRW universes that are asymptotically flat. Regions \RIV\, and \RV\, are asymptotically de Sitter. $\Sigma$ is the conformal 2-sphere defined by the intersection of the lightlike infinity of region \RI\, and the spacelike infinity of region \RIV. The blue curves indicate orbits of the $SO(3,1)$ symmetry, which act as the conformal group on $\Sigma$ \cite{Freivogel:2006xu}. The red lines between regions \RIII, \RIV, and \RV\, represent the cosmological horizons in the dS of the observer at $r=0$. The green curve in region \RIII\, represents the domain wall between the FRW and dS regions.}\label{CDL}
\end{center}
\end{figure}

Instead of $\sim 10^{500}$ vacua let us  follow \cite{Freivogel:2004rd} and  consider a far smaller landscape which possesses only two vacua as a starting point for our construction. This effective potential has only two minima, one corresponding to a positive \CC  and the other  to 0.  In \cite{Freivogel:2004rd}, a $O(D-1)$ symmetric spacetime resulting from the  CDL nucleation process was worked out, the Penrose diagram for it is given in Fig. \ref{CDL}. 
For convenience we reproduce the solution from \cite{Freivogel:2004rd} for $D = 4$, which is the metric for region \RIII\, in Fig. \ref{CDL},
\begin{equation}\label{CDLmet}
ds^{2} = c^{2}dy^{2}+a(y)^{2}\big[d\alpha^2+(\sin^{2}{\alpha})d\beta^{2} - (\sin^{2}{\alpha}\sin^{2}{\beta})d t^{2}\big].
\end{equation}
Here $0\leq y<\pi$, $0\leq\alpha<\pi$, $0\leq\beta < 2\pi$, and $-\infty<t<\infty$; $c$ is a constant that depends on $\Lambda$ 
The solution (\ref{CDLmet}) was obtained by solving the eucliedan CDL equations \footnote{In the setup employed in \cite{Freivogel:2004rd}, the CDL equations for the $O(D-1)$ symmetric euclidean instanton describing the decay of a metastable dS are the standard FRW equations up to changes in sign \cite{Freivogel:2004rd}, 
\begin{align*}
\Big(\frac{\dot{a}}{a}\Big)^{2} &= H^{2} = \frac{8\pi}{3}\Big(\frac{1}{2}\dot{\varphi}^{2} - V(\varphi]\Big)+\frac{1}{a^{2}}\\
\ddot{\varphi}&= -3H\dot{\varphi}+\partial_{\varphi}V[\varphi],
\end{align*} along with the boundary conditions,
\begin{align*}
a\rightarrow c y\hspace{0.25in}(y=0)&\hspace{0.5in}a\rightarrow  c(\pi-y) \hspace{0.25in}(y=\pi)\\
&\partial_{y}\varphi =0\hspace{0.5in}(y=0,\pi).
\end{align*}}. The solution is then continued to Lorentzian signature. The metric for the other regions can be obtained by geodesically completing (\ref{CDLmet}) as is detailed in \cite{Freivogel:2004rd}. The spacetime consists of an asymptotically dS spacetime with an open hyperbolic $\Lambda =0$ FRW bubble inside it. The domain wall (green curve in Fig. \ref{CDL}) is the transition region between the finite $\Lambda$ and $\Lambda =0$ regions; its position and thickness are dependent on specifics of the potential barrier of $V[\varphi]$ \footnote{While we focus on $D=2,\,4$ analysis of the CDL instanton, the framework generalizes if a decompactification occurs in the the transition since the analysis of \cite{Freivogel:2004rd} was for a $O(D-1)$ symmetric solution and we have set $D=4$. The metric structure of (\ref{globaldsmet}), (\ref{FRWtchi}), and (\ref{dsMetconf1}) is the same except $\dOmt$ is replaced be $\dOmd$. The flat region can be also ten or eleven dimensional as the moduli of the compactifcation can role decompactifiying the space.}.  

The analysis is simplified by taking the thin-wall limit \cite{Fabinger:2003gp,Giddings:2004vr,Freivogel:2004rd} ---having the value of the potential barrier's maximum $V_{max}$ large compared to the value of positive minima, i.e., $\Lambda\ll V_{\text{max}}$. This makes the domain wall region sharp and thin. In this limit the solution for $\varphi$ is simplified; outside of the domain wall, $\varphi = \varphi_0$ where the constant $\varphi_0$ is the position of the positive minimum $V[\varphi_0]=\Lambda$ yielding a classical dS region; inside the domain wall (within the  open FRW region), $\varphi$ is at the position of the zero minimum, i.e., $V[\varphi]=0$. Surprisingly there is not a singularity caused by the collapsing FRW geometry as can be seen from the Euclidean geometry. The Lorentzian and Euclidean geometries agree on the spacelike slice in the middle of Fig. \ref{CDL}  and along this slice it is possible to construct a Hartle-Hawking state \cite{1983PhRvD..28.2960H,Freivogel:2004rd,Freivogel:2006xu} to define states for a transition process \footnote{The behavior of the instanton follows the behavior of S-branes in open string field theory \cite{Gutperle:2002ai,FS,Freivogel:2004rd}.}. The position and shape of the domain wall is determined by its tension $\sigma$  which is determined by the width the potential barrier (which is set by the microphysics of the string compactification). For finite $\sigma$ the domain wall is timelike; in taking the limit $\sigma \rightarrow 0$ the throat of the FRW region goes to zero size and the domain wall becomes lightlike; see Fig. \ref{CDLtoconstrainedCDL}.


\section{The transition amplitude}\label{transamp}



The amplitude for the transition is computed as  path integral over all histories that connect the \emph{in} and \emph{out} states, including all possible spacetime configurations, field configurations, as well as configurations of the horizons that would represent the information from the outside multiverse. We must determine the appropriate spacetime region that contains all the information of dS (for example, from a Hartle-Hawking state on a spacelike slice in the middle or \RIII\, of Fig. \ref{CDL}). After picking the gauge choice that a past and future hat are moved to the spacial center of a causal patch; assume that on the spacelike slice in the middle of the center of Fig. \ref{CDL} we construct a Hartle-Hawking state for the spacetime and determine an \emph{out} state. The information within the causal patch is then all that is needed to capture all the information if horizon complementarity is correct. Anything that passes out of the causal patch (goes into region \RIV) will have a complementary description in terms of the highly scrabbled Hawking radiation which will go into region \RI. Therefore region \RI\,  will contain all the information from the Hartle-Hawking state in the middle of region \RIII\, \footnote{This is an operating assumption of the FRW/CFT framework. The actual landscape has $~10^{500}$ vacua leading to region \RIV\, being the rest of the multiverse. The generic scenario has $\mathcal{I}^{+}$  composed of a quantum superposition of fractal multitudes of bubbles possessing different $\Lambda$, interactions, and possibly dimensions \cite{Sekino:2010vc,Bousso:2010pm}, the evolution of which is encoded in the Hawking radiation. The bulk \emph{out} state can then be formulated on a hyperbolic spacelike slice (one of the blue curves in region \RI\, of Fig. \ref{CDL} as well as $\Sigma$)}. In the FRW/CFT framework the spacelike slice is usually taken to be a late time slice, which is an EAdS$_3$ that is dual to the CFT on $\Sigma$ \footnote{It should be noted that we are not constructing the initial and final states on an entire Cauchy 3 surface (for example a blue curve in region I plus the entire future infinity of region IV) as one would do in the initial value problem of GR. This would be an overcounting of the degrees of freedom as we are assuming complementary is in effect.}.

\bwt
\begin{center}
\begin{figure}[h]
\begin{center}
\includegraphics[scale=0.4]{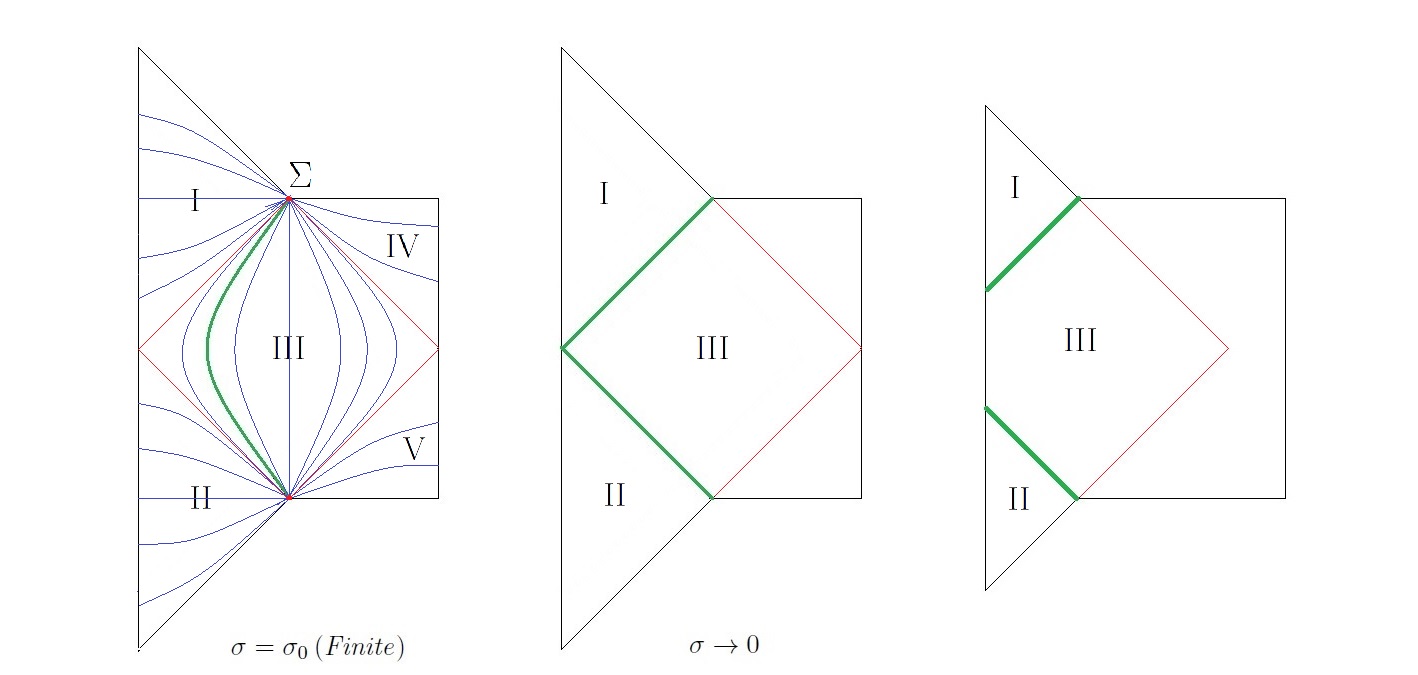}
\end{center}
\begin{center}
\caption{Penrose diagrams of the CDL instanton with finite domain wall tension (left), the limit of the instanton at zero tension (center) and the \emph{constrained} CDL instanton (right).}\label{CDLtoconstrainedCDL}
\end{center}
\end{figure}
\end{center}
\ewt

Let us consider the CDL instanton in this thin-wall tensionless domain wall limit. We  compute a spectral representation of a transition amplitude between \ein and \eout states, $\langle \text{Out}|\text{In}\rangle$, and show that it contains a pole characteristic to a dS intermediate state \cite{goldberger2004collision,2008AIPC.1077...31R}\footnote{We cannot have arbitrary \emph{out} states if we interpret the intermediate region as a metastable dS. One must restrict to \emph{out} states that do not possess enough the energy and entropy to collapse the geometry into a singularity \cite{Banks:2002fe,Bousso:2006ge,Bousso:2010zi}. This state dependence also implies that the evolution of the dS is very dependent on initial and final conditions and in fact most initial states are singular \cite{Hawking27011970}. We are presupposing that the \emph{in} and \emph{out} states are ones that support a singularity free dS. We elaborate more on how this affects our results in the discussion.}
. 



A resonance is an intermediate metastable state that can occur between any initial and final states. Many can be used to establish the existence of a resonance  \footnote{A heuristic example is that many different  states (all those with enough mass energy to form a black hole) can lead to a dense collection of overlapping resonances: a black hole.} and we only need to compute one possible channel that leads to the dS resonance to establish its existence.  A mathematically tractable although not a realistic channel, as it is entropically suppressed, is to construct the \ein and \eout channels in a time-symmetric manner from a semiclassical slice in the middle of region \RIII.

This is not to suggest this channel could be the true cosmological history of our Universe. We are proposing that the existence of a pole in this Rube-Goldberg construction of the channel provides a precise quantum definition in the context of supersymmetric backgrounds of a dS space \footnote{Such time reversal symmetric configurations are employed throughout physics in order to construct resonances \cite{goldberger2004collision,2008AIPC.1077...31R}. As an analogy consider the situation of a black hole in empty Minkowski space that evaporates via hawking radiation. There are many channels (all those with a sufficient amount of matter and radiation) that would create an intermediate resonance (the black hole) that would then Hawking radiate. The spectral representation of any of the channels would possess a dense collection of overlapping poles indicating the presence of the overlapping resonances of a black hole. A time reversal symmetric channel is one where the black hole formed from a cloud of radiation that is essentially identical to the time reverse of the outgoing Hawking radiation. Another more pedestrian analogy would be throwing all the broken bits of a coffee cup together to have them reform the cup, and having the cup bounce up a small distance off the ground, fall back down, and shatter. This extremely entropically suppressed history is by no means the generic history of forming a coffee cup, but can be used to establish the existence of resonance in a transition amplitude associated with a coffee cup.}. This same logic applies to any metastable state in quantum mechanics.

We define the transition amplitude  as a path integral over the histories of the causal patch containing the hats. We do not try to justify this; but study this object's spectral representation and show that it possess a pole that we associate with dS. This eliminates the need to deal with the complicated fractal boundaries, $\mathcal{I}^{+}$ and $\mathcal{I}^{-}$ or regions \RIV\, and \RV.

The full path integral over all histories  contains all fluctuations of the geometry including metric and field configurations about the CDL instanton as well as nonperturbative effects, such as further vacuum decay of the regions outside the hats.  In what follows we  truncate this path integral to only the $\eta_0$ dependence. This minisuperspace approximation focuses the discussion on the first contribution of the transition amplitude, where the only histories that are integrated over are those when no particle content is excited.  The ``off shell" continuation of the CDL instanton in the thin-wall tensionless domain wall limit has the two  FRW regions  with their nucleation points separated by a conformal coordinate time $2\eta_0$. In the limit that $\eta_0\rightarrow 0$ the on shell CDL instanton with zero tension domain wall is restored; see the right diagram of Fig. \ref{CDLtoconstrainedCDL}. This geometry is not a true solution of the CDL equations and has the status of a constrained instanton solution \cite{Frishman:1978xs,Affleck:1980mp,Nielsen:1999vq}. It must be created through cutting and pasting employing the Barrab\'{e}s-Israel junction conditions \cite{PhysRevD.43.1129}, which are demonstrated in Appendix \ref{geom}. We refer to this off shell continuation as the \emph{constrained} CDL instanton. Defining the proper time along the geodesic $\psi=0$ to be $2 t_0$, employing (\ref{globaltimetoconftime}),  we can express the path integral (\ref{calschematic}) as an integration over proper time between the bubbles $ t_0$.

\begin{widetext}
\begin{align}\label{calschematic}
&\langle h_{\text{out}},\varphi_{\text{out}}|h_{\text{in}},\varphi_{\text{in}}\rangle =\int^{h_{\text{out}},\varphi_{\text{out}}}_{h_{\text{in}},\varphi_{\text{in}}}\,\mathcal{D}g\,\mathcal{D}\varphi\,e^{i\,S[g,\varphi]}\sim\mathcal{N}\int\,d\,t_0\,e^{i\,S[t_0]}(1+ \Delta_{fluc.}[\delta g_{\mu,\nu},\delta\varphi] \ldots)\nonumber\\&\hspace{3.75in}+\Delta_{pert}+\text{\emph{instanton/nonperturbative contrib.}}
\end{align}
\end{widetext}
Here $\Delta_{fluc.}[\delta g_{\mu,\nu},\delta\varphi]$ refers to perturbative fluctuations about the \emph{constrained} CDL and $\Delta_{pert}$ refers to all other perturbative ``off shell" history contributions to the path integral. 
This expresses the amplitude as an integral over the relative time between the initial and final hats. The Fourier transform of the $t_0$ dependence defines the spectral representation of $\langle\text{\emph{Out}}|\text{\emph{In}}\rangle$. The terms of the expansion are weighted in powers of $l_{dS}$, which control the expansion.

\begin{figure}[ht]
\begin{center}
\includegraphics[scale=0.4]{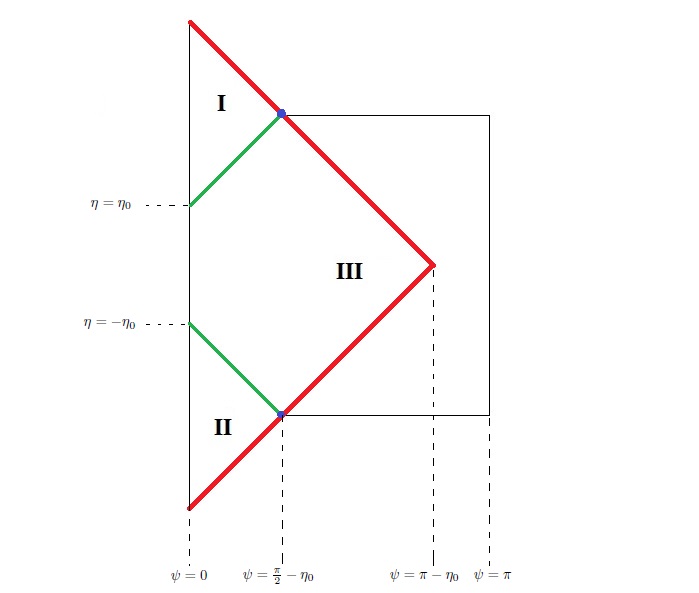}
\end{center}
\caption{Constrained CDL.}\label{constrainedCDL}
\end{figure}

\section{Regulation of the amplitude and \texorpdfstring{$\eta_{0}$}{\texttwosuperior} dependence}\label{calcul}
In the limit and approximations that we are employing, only the $\eta_0$ dependence of the action contributes to the amplitude. In order to compute the action for the causal region of the \emph{constrained} CDL instanton (the region within the red curve of Fig. \ref{constrainedCDL}), we must determine the relevant contributions to the action.

The action contribution of the stress tensor of the domain wall does not depend on $\eta_0$ as can be seen from the boost symmetries of dS. Consider a dS with one hat on $\mathcal{I}^{+}$ that nucleates at a time $\eta_0$ in a particular coordinate frame. Varying the nucleation time, changing $\eta_0$, is equivalent to boosting the frame in the dS. The action contribution of the stress tensor is invariant under these boosts as the action is diffeomorphism invariant. The contribution of the stress tensor is just the stress energy required to change the cosmological constant from $\Lambda$ to 0 as one crosses the domain wall and does not depend on the nucleation time in this limit \footnote{Also within the bubble there is no way to distinguish different nucleation times. The $\eta_0$ dependence of the stress tenor itself found in \ref{geom} is a result of the stress tensor not being diffeomorphism invariant as it is a tensor. The action contribution from this stress tensor does not depend on $\eta_0$.}. Therefore we do not need to include its contribution to the action in the time reversal symmetric amplitude \footnote{An analogy would be to employ intuition from Schwinger pair production, as the effective \CC could be thought of as the electric field and the domain wall as a charge density of pair-produced particles where  field lines end. The bubble is the region of zero electric field between the pair-produced particles. The action contribution of the charge density does not depend on the nucleation time, as the bubble will grow to infinite size and a given nucleation time is simply a gauge choice.}.

The hats of the \emph{constrained} CDL instanton in the approximation that there are no excited particles are described by Milne universes \cite{milne1935relativity}, $ds^{2}_{\text{FRW-Milne}} = -d\tau^2 +\tau^{2}\Big(d\chi^2 + \sinh^{2}{\chi}\dOmt\Big)$, with $0\leq\tau<\infty$ and $0\leq\chi<\infty$. Using the coordinate change $t = \tau\cosh{\chi}$ and $r = \tau\sinh{\chi}$, we can see that this is simply a portion of Minkowski space, the interior of the forward light cone of the origin, $r\leq t$, with hyperbolic slicing.  This means the action contribution of these regions is also $\eta_0$ independent in the limits we are employing; in fact their bulk contributions are semiclassically 0 in the limit of no particles as $R =0$ in this case. 

Therefore we only need to consider the action contribution of the dS region of the causal patch (region \RIII\, of Fig. \ref{constrainedCDL}) in order to get  $\eta_0$ dependence of the transition amplitude in this approximation.

The action of region \RIII\, is divergent due to the infinite volume located at the blue dots in Fig. \ref{constrainedCDL}, and must be properly regulated.  This divergence is present for all values of $\eta_0$ and in all dimensions. The regulator must respect the Lorentz and dS symmetries of the spacetime in order to separate the divergence and $\eta_0$ dependence of the action in an invariant way \footnote{The author thanks Ying Zhao for her many discussions and comments on this point.}. Under boosts and rotations, spacetime points move along surfaces of constant $r^{2}_0 = \frac{3\sin^{2}{\psi}}{\Lambda\cos^{2}{\eta}}$.  For $D>2$, surfaces of constant $r_0$ are those of constant transverse sphere size. When $r_0<\sqrt{\frac{\Lambda}{3}} = l_{dS}$, constant $r_0$ surfaces are timelike and can be identified with the $r$ coordinate of the static patch metric of dS, 
$ds^{2}_{\text{static}}= -\Big(1-\frac{r^{2}}{l^{2}_{dS}}\Big)dt^{2} +\Big(1-\frac{r^{2}}{l^{2}_{dS}}\Big)^{-1}dr^{2} +r^{2}\dOmd $. For $r_0>\sqrt{\frac{\Lambda}{3}}$ the constant $r_0$ surfaces are spacelike and can be identified with the now timelike $r$ coordinate of the future triangle metric, which is identical in form to the static patch metric except $r>l_{dS}$ and is hence timelike \cite{Anninos:2012qw}. The appropriate cutoff procedure is then to restrict the integration region to the portion of region \RIII\, in Fig. \ref{constrainedCDL} that is between the spacelike surface of a fixed given $r_0 >l_{dS}$. \RIII\, is restored in the limit that the cutoff $r_0\rightarrow\infty$. One further regulator is added for convenience here but is necessary in higher dimensions. The two null boundaries of the causal patch intersect in the middle of region \RIII at $\psi =\pi-\eta_0$; we limit the the integration range of $\psi$ to only go to $\psi=\pi-\eta_0-\gamma_0$, with $\gamma_0$ being a small positive constant that avoids the intersection of the null surfaces. In the limit $\gamma_0\rightarrow0$ along with $r_0\rightarrow\infty$ region \RIII is restored. The regulated integration region, $\mathcal{V}$, is then regions enclosed by the red curves in Fig. \ref{spacetime1p1} in $1+1$ dimensions  and Fig. \ref{spacetime} in higher dimensions.

\section{The \texorpdfstring{$1+1$}{1+1\texttwosuperior} dimensional Action in Liouville Gravity}\label{onepone}
We  first compute the amplitude in the context of dS$_2$.  This can be described by Lorentzian timelike Liouville gravity \cite{Polchinski:1989fn,Seiberg:1990eb,Polchinski:1990mh,Ginsparg:1993is,Dorn:1994xn,Zamolodchikov:1995aa,Zamolodchikov:2005fy} which contains dS$_{2}$ as a solution \cite{Ambjorn:1998fd,Loll:1999uu,Harlow:2011ny}\footnote{Here the timelike Liouville action is defined and evaluated via analytic continuation as shown in \cite{Harlow:2011ny} with the remaining $SL_{2}\mathbb{C}$ gauge  redundancy of topologically simple spacetimes fixed by the procedure of \cite{Maltz:2012zs}. $\mu$ is the Liouville cosmological constant and the ratio $\Lambda/\mu$ should be thought of as expressing $\Lambda$ in terms of the UV Planck scale.}.  This dramatically simplifies the calculation since in $1+1$ dimensions the Gauss-bonnet theorem implies that the contribution of the boundary of $\mathcal{V}$ integrates to $\mathcal{V}$'s Euler characteristic and is $\eta_0$ independent. In $1+1$ dimensions we therefore only need to consider the bulk contributions of the action.  

\begin{figure}[h]
\begin{center}
\includegraphics[scale=0.4]{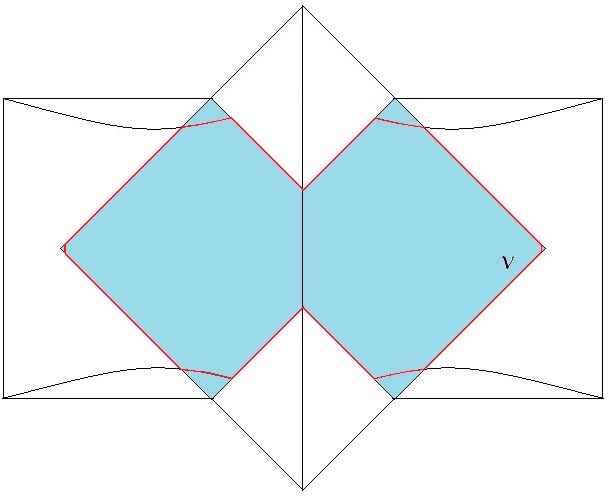}
\end{center}
\caption{Penrose diagram of the $1+1$ dimensions \emph{constrained} instanton with the integration region shaded in blue. The slices of constant  $r^{2}_0 = \frac{3\sin^{2}{\psi}}{\Lambda\cos^{2}{\eta}}$ are the curved surfaces intersecting the null lines at $\psi_{1}=\arctan{\Bigg[\frac{\cos{[\eta_{0}]}}{\sin{\eta_{0}}+\sqrt{\frac{3}{\Lambda r^{2}_{0}}}}\Bigg]}$ and $\psi_{2}=\frac{\pi}{2}-\arctan{\Bigg[\frac{\sin{\eta_{0}}-\sqrt{\frac{3}{\Lambda r^{2}_{0}}}}{\cos{\eta_0}}\Bigg]}$ as well as their reflection about $\psi=0$. The null domain walls dividing the dS and FRW regions intersect $\psi=0$ at conformal time $\eta=\eta_0$ and $\eta=-\eta_0$. The regulated integration region, $\mathcal{V}$, is the volume enclosed by the red curve. Taking the cutoff $r_0\rightarrow\infty$ restores the entire integration region. }\label{spacetime1p1}
\end{figure}

The timelike Liouville action is then

\begin{equation}\label{oplusoaction}
S_{L}=-\frac{1}{16\pi b^{2}}\int_{\mathcal{V}}\,d^{2}\xi\big(\eta^{ab}\partial_{a}\phi_c\partial_{b}\phi_c -16\lambda e^{\phi_c}\big).
\end{equation}

Here the  metric is put into conformal gauge \cite{Ginsparg:1993is,Nakayama:2004vk,Harlow:2011ny,Maltz:2012zs} $g_{ab}=e^{\phi_c}\eta_{ab}$ and $e^{\phi_c}=\frac{3}{\Lambda\cos^{2}{\eta}}$.
Via the Liouville equation of motion we have
\begin{equation}\label{lioueom}
\frac{1}{4}\eta^{ab}\partial_{a}\partial_{b}\phi_c = -2\lambda e^{\phi_c} = -\frac{2\cdot 3\cdot\lambda}{\Lambda\cos^{2}{\eta}},
\end{equation}
which gives $\lambda=\frac{\Lambda}{3\cdot 4}$.

In $D$ spacetime dimensions the  regulated boundary ---red curve in Figs. \ref{spacetime1p1} and \ref{spacetime}--- is the surface described by the following curves times the transverse $\mathbb{S}^{D-2}$,

\begin{align}\label{regionv}
\eta_{1}&=\psi+\eta_{0}\hspace{0.95in}\psi\in[0,\psi_{1}]\\
\eta_{2}&=\arccos{\Bigg[\sqrt{\frac{3}{\Lambda}}\frac{\sin{\psi}}{r_{0}}\Bigg]}\hspace{0.2in}\psi\in[\psi_{1},\psi_{2}]\\
\eta_{3}&=\pi-(\psi+\eta_{0})\hspace{0.6in}\psi\in[\psi_2,\pi-\eta_{0}-\gamma_{0}]\\
\psi_{4}&=\pi-\gamma_{0}-\eta_{0}\hspace{0.68in}\eta\in[-\gamma_0,\gamma_0]\\
\eta_{5}&=-(\psi+\eta_{0})\hspace{0.745in}\psi\in[0,\psi_{1}]\\
\eta_{6}&=-\arccos{\Bigg[\sqrt{\frac{3}{\Lambda}}\frac{\sin{\psi}}{r_{0}}\Bigg]}\hspace{0.075in}\psi\in[\psi_{1},\psi_{2}]\\
\eta_{7}&=(\psi+\eta_{0})-\pi\hspace{0.585in}\psi\in[\psi_2,\pi-\eta{0}-\gamma_{0}]\label{regionvf}
\end{align}

Here $\psi_1 = \arctan{\Bigg[\frac{\cos{\eta_{0}}}{\sin{\eta_{0}}+\sqrt{\frac{3}{\Lambda r^{2}_{0}}}}\Bigg]}$ and $\psi_2 = \frac{\pi}{2}- \arctan{\Bigg[\frac{\sin{\eta_{0}}-\sqrt{\frac{3}{\Lambda r^{2}_{0}}}}{\cos{\eta_{0}}}\Bigg]}$ are where the constant $r_0$ surfaces intersect the null boundaries. In $1+1$ dimensions the transverse sphere is an $\mathbb{S}^{0}$, which is just two points, leading to the Penrose diagram in Fig. \ref{spacetime1p1} 
Therefore $\mathcal{V}$ in $1+1$ dimensions is the region enclosed by (\ref{regionv})-(\ref{regionvf}) and its reflection across $\psi=0$. Inserting this into (\ref{oplusoaction}) we have

\bwt
\begin{align}\label{liouint}
S_{L}&=\frac{4}{16\pi b^{2}}\int_{\mathcal{V}}\,d\psi\,d\eta\Bigg(\frac{1+\sin^{2}{\eta}}{\cos^{2}{\eta}}\Bigg)=\frac{1}{\pi b^{2}}\Bigg\{\int^{\psi_1}_{0}d\psi\int^{\psi+\eta_0}_{0}d\eta\Bigg(\frac{1+\sin^{2}{\eta}}{\cos^{2}{\eta}}\Bigg)\nonumber\\&\hspace{0.5in}+\int^{\psi_2}_{\psi_1}d\psi\int^{\arccos{\big[\sqrt{\frac{3}{\Lambda}}\frac{\sin{\psi}}{r_0}}\big]}_{0}d\eta\,\frac{1+\sin^{2}{\eta}}{\cos^{2}{\eta}}+\int^{\pi-\eta_0-\gamma_0}_{\psi_2}d\psi\int^{\pi-(\psi+\eta_0)}_{0}d\eta\Bigg(\frac{1+\sin^{2}{\eta}}{\cos^{2}{\eta}}\Bigg)\Bigg\}.
\end{align}
\ewt

After preforming the $\eta$ integration in all three terms of (\ref{liouint}), we see that the integrand resulting from the second term in (\ref{liouint}) is bounded within its $\psi$ integration range. In the cutoff limit $r_0\rightarrow\infty$, $\psi_{1}\rightarrow\psi_{2}$, therefore the middle integral goes to 0 in the limit and can be ignored. 

After computing (\ref{liouint}) and taking the $\gamma_0\rightarrow0$ limit we can  Laurent expand (\ref{liouint}) in $w_0=1/r_0$ up to $O[w_0]$ resulting in

\begin{equation}
S_{L}=-\frac{1}{\pi b^{2}}\Bigg\{4\log{\Bigg|\sqrt{\frac{3}{\Lambda}}w_0\Bigg|}-4+\frac{\pi^{2}}{4}-\frac{\eta^{2}_{0}}{2}+2\log{\cos{\eta_0}}\Bigg\}.\nonumber
\end{equation}

The $-\frac{4!\mu}{2\Lambda}\big\{\log{\Big|\sqrt{\frac{3}{\Lambda}}w_0\Big|}-4+\frac{\pi^{2}}{4}\big\}$ term is the divergent contribution of the action that remains when $\eta_0=0$. This divergence, resulting from the infinite volume of region \RIII, was to be expected and is just the action of the Lorentizan tensionless domain-wall CDL instanton, $S_0$, in this limit. When exponentiated it can be absorbed into the overall normalization factor of (\ref{calschematic}).

Defining $\tilde{S}_{L} = S_L - S_0$ and reexpressing this in terms of proper time $t_0$ using (\ref{globaltimetoconftime}) results in \footnote{$\lambda =\pi\mu b^{2}$ so $\frac{1}{2\pi b^{2}}=\frac{\pi\mu}{2\pi\lambda}=\frac{4!\mu}{4\Lambda}$}

\bwt
\begin{align}\label{1plus1globtime}
\tilde{S}_{L}\hspace{-1mm}&=\hspace{-1mm}\frac{2}{\pi b^{2}}\Bigg\{\arctan^{2}{\Bigg[\tanh{\Bigg[\sqrt{\frac{\Lambda}{3}}t_0\Bigg]}\Bigg]}\hspace{-1mm}+\hspace{-1mm}\log{\cosh{\Bigg[\sqrt{\frac{\Lambda}{3}}t_0\Bigg]}}\Bigg\}=\frac{4!\mu}{4\Lambda}\Bigg\{\arctan^{2}{\Bigg[\tanh{\Bigg[\sqrt{\frac{\Lambda}{3}}t_0\Bigg]}\Bigg]}\hspace{-1mm}+\hspace{-0.5mm}\log{\cosh{\Bigg[\sqrt{\hspace{-1mm}\frac{\Lambda}{3}}t_0\hspace{-0.25mm}\Bigg]}}\Bigg\}.
\end{align}
\ewt

The $\log{\cosh{\Big[\sqrt{\frac{\Lambda}{3}}t_{0}\Big]}}=\sqrt{\frac{\Lambda}{3}}t_{0}+\log{|1+e^{-2\sqrt{\frac{\Lambda}{3}}t_0}|}-\log{2}$ term in (\ref{1plus1globtime}) is the only $t_0$ dependent term that is not bounded. We see that for large values of $t_0$ the action grows linearly with $t_0$.

Treating the bounded term as a perturbation and Fourier transforming with respect to $t_0$ yields,

\bwt
\begin{align}\label{1p1pole}
\int^{\infty}_{0}\hspace{-1mm}d\,t_{0}\,e^{i (\tilde{S}_L[t_0] - \omega\,t_0)}&=\int^{\infty}_{0\hspace{-1mm}}d\,t_{0}\, e^{i(2\mu\sqrt{\frac{3}{\lambda}}t_{0}-\omega t_0)}\Bigg(1+i\frac{3\cdot 2\mu}{\Lambda}\Bigg\{\log{\Big|\frac{1+e^{-2\sqrt{\frac{\Lambda}{3}}t_{0}}}{2}\Big|}+\arctan^{2}{\hspace{-1mm}\Bigg[\tanh{\sqrt{\frac{\Lambda}{3}t_{0}}}\Bigg]}\Bigg\}+\ldots\Bigg)\nonumber\\&= \frac{i}{\omega -2\mu\sqrt{\frac{3}{\Lambda}}} + \rho_{1}[\omega] +\ldots.
\end{align}
\ewt
thus revealing a pole in the spectral representation. One notes  that $2\mu\sqrt{\frac{3}{\Lambda}}$ is the energy of the static patch of dS, we take the existence of this pole to be the indication of an intermediate dS vacuum. 

This indicates that the dS can be thought of as a resonance in a transition amplitude. 

The pole in (\ref{1p1pole}) occurs at a real value of $\omega$ but this is an approximation. When the metastable character of the dS vacuum is accounted for the cosmological constant obtains a small imaginary part determined by the CDL decay rate. This shifts the pole by a slightly imaginary amount, which is standard in the analysis of resonances \cite{0201503972,2008AIPC.1077...31R}.

$\rho_{1}[\omega]$ is the contribution of the $O\big[\frac{\mu}{\Lambda}\big]$ term in (\ref{1p1pole}). The first term which can be integrated employing  ${}_{2}F_{1}$ hypergeometric functions and the Lebesgue dominant convergence theorem; the second term is a bounded function of $t_0$ and gives further contributions to the spectral representation along with the rest of the expansion.

\section{The amplitude computation in the context of \texorpdfstring{$3+1$}{3+1\texttwosuperior} dimensions}\label{threepone}

\begin{figure}[t]
\begin{center}
\includegraphics[scale=0.4]{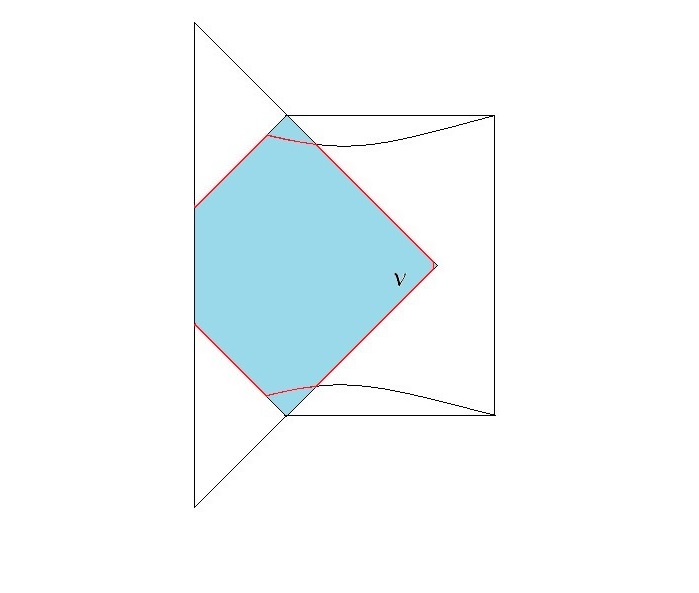}
\end{center}
\caption{$\mathcal{V}$ for the $d+1$ spacetime.}\label{spacetime}
\end{figure}
Now that we have established that the spectral representation of the $1+1$ dimensional amplitude possesses poles associated with dS, let us repeat this in $3+1$ dimensions  in GR limit. Again we employ the cutoff region to be $\mathcal{V}$ with the $r^{2}_0 = \frac{3\sin^{2}{\psi}}{\Lambda\cos^{2}{\eta}}$, which respects the lorentz and dS symmetries. Here surfaces of constant $r_0$ are surfaces of constant transverse $\mathbb{S}^{2}$. This implies that the region of integration is $\mathcal{V}$, which is the red curve in Fig. \ref{spacetime}. In order to properly compute the action we must include the boundary contributions. Therefore we must append to the Einstein-Hilbert action \cite{Einstein:1916cd}, the Gibbons-Hawking-York (GHY) spacelike boundary term \cite{PhysRevLett.28.1082,PhysRevD.15.2752}, its null generalization \cite{Parattu:2015gga,Parattu:2016trq}, and the contribution of corner terms \cite{PhysRevD.47.3275,Brown:2015lvg,Lehner:2016vdi}.  This leads to the action 

\begin{widetext}
\begin{equation}\label{actionmaster}
S=\frac{1}{2\kappa}\int_{\mathcal{V}}\,d^{4}x\sqrt{-g}\big(R-2\Lambda\big)-\sum_{i=2,4,6}\frac{1}{2\kappa}\int_{\partial\mathcal{V}_{i}}\,d^{3}x 2\sqrt{h_{(i)}}K_{(i)}+\sum_{i=1,3,5,7}\frac{1}{2\kappa}\int_{\partial\mathcal{V}_{i}}\,d^{2}x\sqrt{q_{(i)}}\mathit{\Theta} +\sum^{5}_{j=1}S_{\text{corner},(j)}.
\end{equation}
\end{widetext}
Here the GHY term is composed of the extrinsic curvature $K_{ab}= e^{\mu}_{a}e^{\nu}_{b}\nabla_{\nu}n_{\mu,(i)}$ with $i =2,4,6$ referring to the normals, (\ref{normals2}), (\ref{normals4}), and (\ref{normals6}), and $h_{ab}$ is the intrinsic metric on the boundary. On the null boundaries $i=1,3,5,7$ with normals (\ref{normals1}), (\ref{normals3}), (\ref{normals5}), and (\ref{normals7}), the null generalization consists of the metric of the transverse $\mathbb{S}^{2}$, $q_{AB}$, and the second fundamental form on the null surface $\mathit{\Theta}_{ab} = q^{c}_{a}q^{d}_{b}\nabla_{c}l_{d}$, with resulting scalar $\mathit{\Theta} = q^{ab}\mathit{\Theta_{ab}} = \frac{1}{2}q^{AB}\mathcal{L}_{l}q_{AB} =\frac{1}{\sqrt{q}}\frac{d}{d\psi}\sqrt{q}$     following the conventions of \cite{Parattu:2015gga,Parattu:2016trq}.


\subsection{Bulk action}
The bulk integration in $\mathcal{V}$ is the region bounded by the surfaces in (\ref{regionv})-(\ref{regionvf}); this makes the bulk action contribution,
\bwt
\begin{align}\label{Bulkactionint}
S_{Bulk}&=\frac{1}{2\kappa}\int_{\mathcal{V}}\,d^{4}x\sqrt{-g}\big(R-2\Lambda\big)=\frac{2\cdot 4\pi\cdot 2\Lambda}{2\kappa}\Bigg\{\int^{\psi_1}_{0}\,d\psi\int^{\psi+\eta_0}_{0}d\eta\Big(\frac{3}{\Lambda}\Big)^{2}\frac{\sin^{2}{\psi}}{\cos^{4}{\eta}}\nonumber\\&\hspace{0.75in}+\int^{\psi_2}_{\psi_1}\,d\psi\int^{\arccos{\Big[\sqrt{\frac{3}{\Lambda r^{2}_{0}}}\sin{\psi}\Big]}}_{0}\,d\eta\Big(\frac{3}{\Lambda}\Big)^{2}\frac{\sin^{2}{\psi}}{\cos^{4}{\eta}}+\int^{\pi -\eta_0-\gamma_0}_{\psi_2}\,d\psi\int^{\pi-(\psi+\eta_0)}_{0}\,d\eta\Big(\frac{3}{\Lambda}\Big)^{2}\frac{\sin^{2}{\psi}}{\cos^{4}{\eta}}\Bigg\}.
\end{align}
\ewt

When integrated this yields

\bwt
\begin{align}\label{Bulkactionlimit}
S_{Bulk} &= \frac{4\pi4!}{2\kappa\Lambda}\Bigg\{\Bigg(2-2\log{\big|\cos{\eta_0+\psi}\big|}+\frac{\sin^{2}{\eta_0}}{\cos^{2}{[\psi+\eta_0]}}-\cos{[2\psi]}\tan^{2}{[\psi+\eta_0]}\Bigg)\Bigg|^{\psi_1}_{0}\nonumber\\&\hspace{0.35in}+\frac{1}{12\sqrt{3}}\Bigg(-2r^{3}_{0}\Lambda^{3/2}\textit{arctanh}{\Bigg[\frac{\sqrt{2}\Lambda r_{0}\cos{\psi}}{\sqrt{2 \Lambda r^{2}_{0} +3\cos{[2\psi]}-3}}\Bigg]}-3\cos{\psi}\sqrt{4r^2_{0}\Lambda +6\cos{[2\psi]}-6}\nonumber\\&\hspace{0.75in}+6\sqrt{3}\log{\Bigg|\frac{\sqrt{6}\cos{\psi}+\sqrt{2r^{2}_0\Lambda+3\cos{[2\psi]}-3}}{\sqrt{\Lambda}r_0}\Bigg|}\Bigg)\Bigg|^{\psi_2}_{\psi_1}+\frac{1}{4}\Bigg(-2+2\log{\big|\cos{\eta_0+\psi}\big|}\nonumber\\&\hspace{1.15in}-\frac{\sin^{2}{\eta_0}}{\cos^{2}{[\psi+\eta_0]}}+\cos{[2\psi]}\tan^{2}{[\psi+\eta_0]} \Bigg)\Bigg|^{\pi-\eta_0-\gamma_0}_{\psi_2}\Bigg\}.
\end{align}
\ewt
Taking the cutoff limit $r_0\rightarrow\infty$ and $\gamma_0\rightarrow 0$ makes $\mathcal{V}$ into region \RIII. If we express (\ref{Bulkactionlimit}) as a Laurent expansion of $w_0=\frac{1}{r_0}$ after taking the $\gamma_0\rightarrow 0$, combining terms, and exploiting trigonometric identities, we finally get to $O[\omega_0]$,
\begin{align}\label{bulkactionlimita}
S_{Bulk}&=\frac{4\pi4!}{2\kappa\Lambda}\Bigg\{\frac{\Lambda}{2w^{2}_0}+\frac{1}{2}\log{\frac{\Lambda}{3w^{2}_0}} +\frac{5}{24}\nonumber\\&\hspace{0.75in}+\frac{1}{8}\cos{[2\eta_0]} -\frac{1}{2}\log{|\cos{\eta_0}|}\Bigg\}.
\end{align}

\subsection{Boundary action}

The boundary contributions of the action (\ref{actionmaster}) depend on the normals of the boundaries detailed in (\ref{normals1})-(\ref{normals7}).

The outward \footnote{This convention is different from the one adopted in section \ref{geom}, where, for the junction conditions we used future directed normals as opposed to outward directed normals.} directed normal one-forms and their associated vectors are
\bwt
\begin{align}
&n_{(1)\alpha}=\big(\delta^{\eta}_{\alpha}-\delta^{\psi}_{\alpha}\big)\hspace{1in}n^{\alpha}_{(1)}=-\frac{\Lambda\cos^{2}{(\psi+\eta_0)}}{3}\big(\delta^{\alpha}_{\eta}+\delta^{\alpha}_{\psi}\big)\label{normals1}\\&n_{(2)\alpha}=\sqrt{\frac{3}{\Lambda}}\frac{\sqrt{\Lambda r^{2}_{0}-3\sin^{2}{\psi}}}{\cos{\eta}\sqrt{\Lambda r^{2}_{0}-3}}\big(\delta^{\alpha}_{\eta}+\frac{\sqrt{3}\cos{\psi}}{\sqrt{\Lambda r^{2}_{0}-3\sin^{2}{\psi}}}\delta^{\alpha}_{\psi}\big)\nonumber\\&n^{\alpha}_{(2)}=\sqrt{\frac{\Lambda}{3}}\frac{\sqrt{\Lambda r^{2}_{0}-3\sin^{2}{\psi}}\cos{\eta}}{\sqrt{\Lambda r^{2}_{0}-3}}\big(-\delta^{\alpha}_{\eta}+\frac{\sqrt{3}\cos{\psi}}{\sqrt{\Lambda r^{2}_{0}-3\sin^{2}{\psi}}}\delta^{\alpha}_{\psi}\big)\label{normals2}\\&n_{(3)\alpha}=\big(\delta^{\eta}_{\alpha}+\delta^{\psi}_{\alpha}\big)\hspace{1in}n^{\alpha}_{(3)}=\frac{\Lambda\cos^{2}{(\psi+\eta_0)}}{3}\big(-\delta^{\alpha}_{\eta}+\delta^{\alpha}_{\psi}\big)\label{normals3}\\&n_{(4)\alpha}=\sqrt{\frac{3}{\Lambda}}\frac{1}{\cos{\eta}}\delta^{\psi}_{\alpha}\hspace{0.85in}n^{\alpha}_{(4)}=\sqrt{\frac{\Lambda}{3}}\cos{\eta}\,\delta^{\alpha}_{\psi}\label{normals4}\\&n_{(5)\alpha}=\big(-\delta^{\eta}_{\alpha}+\delta^{\psi}_{\alpha}\big)\hspace{1in}n^{\alpha}_{(5)}=\frac{\Lambda\cos^{2}{(\psi+\eta_0)}}{3}\big(\delta^{\alpha}_{\eta}+\delta^{\alpha}_{\psi}\big)\label{normals5}\\&n_{(6)\alpha}=\sqrt{\frac{3}{\Lambda}}\frac{\sqrt{\Lambda r^{2}_{0}-3\sin^{2}{\psi}}}{\cos{\eta}\sqrt{\Lambda r^{2}_{0}-3}}\big(-\delta^{\alpha}_{\eta}+\frac{\sqrt{3}\cos{\psi}}{\sqrt{\Lambda r^{2}_{0}-3\sin^{2}{\psi}}}\delta^{\alpha}_{\psi}\big)\nonumber\\&n^{\alpha}_{(6)}=\sqrt{\frac{\Lambda}{3}}\frac{\sqrt{\Lambda r^{2}_{0}-3\sin^{2}{\psi}}\cos{\eta}}{\sqrt{\Lambda r^{2}_{0}-3}}\big(\delta^{\alpha}_{\eta}+\frac{\sqrt{3}\cos{\psi}}{\sqrt{\Lambda r^{2}_{0}-3\sin^{2}{\psi}}}\delta^{\alpha}_{\psi}\big)\label{normals6}\\&n_{(7)\alpha}=\big(-\delta^{\eta}_{\alpha}-\delta^{\psi}_{\alpha}\big)\hspace{1in}n^{\alpha}_{(7)}=\frac{\Lambda\cos^{2}{(\psi+\eta_0)}}{3}\big(\delta^{\alpha}_{\eta}-\delta^{\alpha}_{\psi}\big).\label{normals7}
\end{align}
\ewt

With the scalar extrinsic curvature defined as $K= -\nabla_{\alpha}n^{\alpha}$

we have
\begin{align}
\mathit{\Theta}_{(1)}&=\frac{1}{\sqrt{q_{(1)}}}\frac{d}{d\psi}\sqrt{q_{(1)}}\nonumber\\
K_{(2)}&=\frac{\cos{\psi}\sqrt{6\cos{[2\psi]}+4\Lambda r^{2}_{0}-6}\hspace{-0.4mm}-\hspace{-0.4mm}3\sqrt{3}\cos{[2\psi]}\hspace{-0.4mm}-\hspace{-0.4mm}\sqrt{3}}{2r_{0}\sqrt{\Lambda r^{2}_{0}-3}}\nonumber\\
\mathit{\Theta}_{(3)}&=\frac{1}{\sqrt{q_{(3)}}}\frac{d}{d\psi}\sqrt{q_{(3)}}\nonumber\\
K_{(4)}&=-2\sqrt{\frac{\Lambda}{3}}\cot{[\pi-\eta_{0}-\gamma_{0}]}\cos{\eta}\nonumber
\end{align}

This make the boundary action

\begin{align}
S_{Boundary}&= -\sum_{i=2,4,6}\frac{1}{2\kappa}\int_{\partial\mathcal{V}_{i}}\,d^{3}x 2\sqrt{h_{(i)}}K_{(i)}\nonumber\\&\hspace{0.25in}+\sum_{i=1,3,5,7}\frac{1}{2\kappa}\int_{\partial\mathcal{V}_{i}}\,d^{2}x\sqrt{q_{(i)}}\mathit{\Theta}_{(i)}
\end{align}

equal to
\bwt
\begin{align}\label{boundaction}
S_{Boundary}&=\frac{4\cdot4\pi}{2\kappa}\Bigg\{\int^{\psi_1}_{0}\frac{d\psi}{2}\frac{\partial}{\partial\psi}\Bigg(\frac{3\sin^{2}{\psi}}{\Lambda\cos^{2}{[\psi+\eta_0]}}\Bigg)-\int^{\psi_2}_{\psi{1}}\Bigg(\frac{3\sin^{3}{\psi}}{\Lambda\cos^{2}{\big[\sqrt{\frac{3}{\Lambda}}\frac{\sin{\psi}}{r_0}\big]}}\Bigg)^{3/2}K_{(2)}\nonumber\\&\hspace{0.55in}+\int^{\pi-\eta_0-\gamma_0}_{\psi_2}\frac{d\psi}{2}\frac{\partial}{\partial\psi}\Bigg(\frac{3\sin^{2}{\psi}}{\Lambda\cos^{2}{[\psi+\eta_0]}}\Bigg)+\int^{0}_{\gamma_0}\,d\eta\Bigg(\frac{3}{\Lambda\cos^{2}{\eta}}\Bigg)^{3/2}\sin^{2}{[\pi-\eta_0-\gamma_0]}K_{(4)}\Bigg\}.
\end{align}
\ewt

In the cutoff limit $r_0\rightarrow\infty$, $\psi_1\rightarrow\psi_2$, and $K_{(2)}\rightarrow 0$. Therefore, the second term in (\ref{boundaction}) does not contribute.

Upon integration (\ref{boundaction}) yields

\begin{align}\label{boundaryactionlimit}
S_{Boundary}&=\frac{4!4\pi}{2\kappa\Lambda}\Bigg\{\frac{\Lambda r^{2}_{0}}{3\cdot 2\cdot 2}+\frac{3\Lambda\sin^{2}{\eta_0}}{2\cdot 2\cdot 3\Lambda} -\frac{\Lambda r^{2}_{0}}{2\cdot 2\cdot 3}\nonumber\\&\hspace{0.75in} -\frac{3\Lambda\cdot 2}{3\cdot 2\Lambda}\log{4}\sin^{2}{\eta_0}\Bigg\}\nonumber\\&=\frac{4!4\pi}{2\kappa\Lambda}\Big\{\frac{1}{4}-\log{4}\Big\}\sin^{2}{\eta_0}.
\end{align}

\subsection{Corner terms}

\begin{figure}[t]
\begin{center}
\includegraphics[scale=0.8]{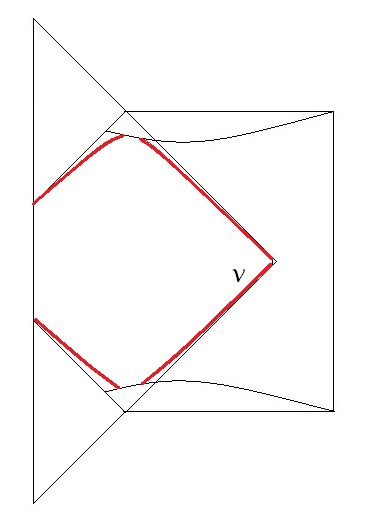}
\end{center}
\caption{The four divergent corner terms that occur are independent of $\eta_0$. The corner contribution is dependent on the boost angle and $\mathbb{S}^{2}$ area, neither of which depend on $\eta_0$. This can be seen as $r_0$ can be varied independently of $\eta_0$, implying that the $\mathbb{S}^{2}$ area is independent of $\eta_0$. The boost angle is also $\eta_0$ independent; this can be seen by deforming the boundaries of $\mathcal{V}$ to spacelike curves (red curves) which intersect the $r_0$ surface at finite $\eta_0$ independent boost angle. In the limit that this spacelike parametrically becomes null the integration region $\mathcal{V}$ is restored.}\label{cornercutoff}
\end{figure}

\begin{figure}[ht]
\begin{center}
\includegraphics[scale=0.4]{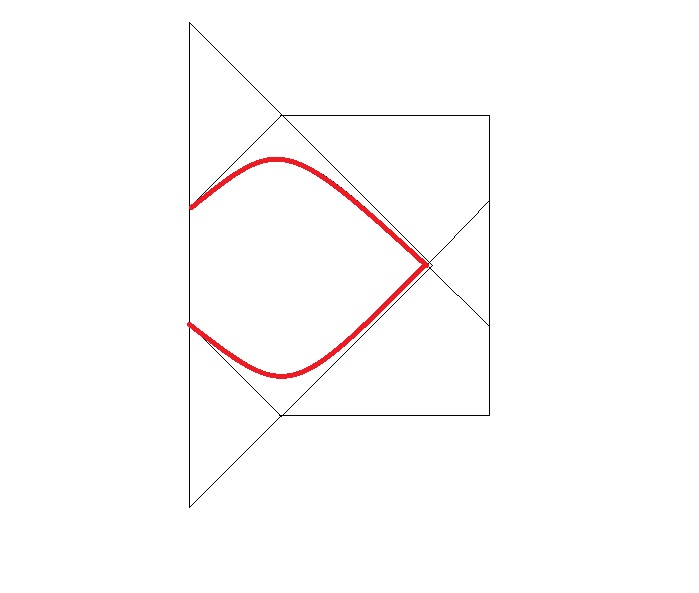}
\end{center}
\caption{If the integration region $\mathcal{V}$ is deformed to the spacelike surfaces, red curves, the divergence of the remaining corner term can be analyzed. In $2+1$ dimensions the Wick rotation of this analysis was carried out in \cite{Brown:2015bva,Brown:2015lvg}.}\label{spacetimeying}
\end{figure}

Finally we must speak of the contributions of the corner terms. I  argue that with the exception of the corner term on the waist of the dS hyperboloid, the action contributions of corner terms are independent of $\eta_0$. The action contribution of the corner resulting from two intersecting hypersurfaces depends on the boost angle and the area of the $\mathbb{S}^2$ at the intersection point \cite{PhysRevD.47.3275,Lehner:2016vdi}.  In our setup there are six corner contributions: the intersection of the constant $r_0$ surface with the null walls at $\psi_1$ and $\psi_2$ and two at the waist. For the four nonwaist contributions the corner is on the curve of constant $\mathbb{S}^2$ radius $r_0$, which is independent of $\eta_0$ ($r_0$ can be varied without changing $\eta_0$). The boost angle at these four points while infinite is independent of $\eta_0$; this can be seen by treating the null surface as the limit of a sequence of spacelike surfaces that emanate from the nucleation point of the respective hat and intersect the constant $r_0$ surface at a point in between $\psi_1$ and $\psi_2$; see Fig. \ref{cornercutoff}. The boost angle for this corner term is finite and is independent of $\eta_0$ as the intersection point can be varied without moving $\eta_0$. In the limit that the spacelike surfaces become null, the corner contributions become infinite but remain $\eta_0$ independent and can be absorbed in the divergent $\eta_0$ independent action term that comes from the original CDL instanton. Hence the only troublesome point is the corner terms at the waist; which have infinite boost angles times an $\eta_0$ dependent finite $\mathbb{S}^2$ size. For paths close to the CDL instanton, $t_0\rightarrow0$, this term vanishes exponentially. In the large $t_0$ limit the derivative of this term with respect to $t_0$ goes to 0 implying that this term becomes constant in the large $t_0$ limit. This term is not well understood and relates to the specification of microstates of the horizon and requires a better understanding of the horizon degrees of freedom perhaps employing some stretched horizon analysis. The calculation employed here in $2+1$ dimensions is closely related to Wick rotations of those in \cite{Brown:2015bva,Brown:2015lvg,PhysRevD.47.3275,Lehner:2016vdi}, which relate complexity and action. This term also appears in their analysis of the null and corner terms, and an analysis of it was carried out employing the spacelike cutoffs in Fig. \ref{spacetimeying}.

\section{total action and the pole}\label{totsactionandpole}

Combining  the terms  (\ref{bulkactionlimita}) and (\ref{boundaryactionlimit}) we have the total action, which after Laurent expanding in $w_0 =\frac{1}{r_0}$  up to $O[w_0]$ results in

\begin{align}
S&=\frac{4\pi4!}{2\kappa\Lambda}\Bigg\{-\frac{1}{2}\log{|\cos{\eta_0}|}+\Big\{\frac{1}{4}-\log{4}\Big\}\sin^{2}{\eta_0}\nonumber\\&\hspace{0.15in}+\frac{1}{8}\cos{[2\eta_0]}+\frac{\Lambda}{2w^{2}_0}+\frac{1}{2}\log{\frac{\Lambda}{3w^{2}_0}} +\frac{5}{24} \Bigg\} + S_{\text{corner}};\nonumber
\end{align}
reexpressing $S$  in terms of the proper time $t_0$ using (\ref{globaltimetoconftime}) and renaming the divergent $t_0$ independent constant in (\ref{totsactiont}) to $S_0$, we can define $\tilde{S} = S - S_{0}$ resulting in

\begin{align}\label{totsactiont}
\tilde{S}&=\frac{4\pi4!}{2\kappa\Lambda}\Bigg\{\frac{1}{8}\Bigg(\frac{1-\sinh^{2}{\Big[\sqrt{\frac{\Lambda}{3}}t_0\Big]}}{\cosh^{2}{\Big[\sqrt{\frac{\Lambda}{3}}t_0\Big]}}\Bigg)\nonumber\\&\hspace{0.2in}+\frac{1}{2}\log{\Bigg|\cosh{\Bigg[\sqrt{\frac{\Lambda}{3}}t_0\Bigg]}\Bigg|}\nonumber\\&\hspace{0.4in}+\Bigg\{\frac{1}{4}-\log{4}\Bigg\}\tanh^{2}{\Bigg[\sqrt{\frac{\Lambda}{3}}t_0\Bigg]}\Bigg\},
\end{align}

with $S_{0} = \frac{4\pi4!}{2\kappa\Lambda}\Big\{\frac{\Lambda}{2w^{2}_0}+\frac{1}{2}\log{\frac{\Lambda}{3w^{2}_0}} +\frac{5}{24}\Big\}+S_{\text{corner}}$. 
Apart from the $\log{\cosh{\Big[\sqrt{\frac{\Lambda}{3}t_0}\Big]}}$ term the $t_0$ dependent terms of (\ref{totsactiont}) are bounded and monotonic for $t_0>0$. 
\begin{align}
\tilde{S}&=\frac{4\pi4!}{2\kappa\Lambda}\Bigg\{\frac{1}{2}\log{|\cosh{\Big[\sqrt{\frac{\Lambda}{3}}t_{0}}\Big]|}+\frac{1-\sinh^{2}{\Big[\sqrt{\frac{\Lambda}{3}}t_0\Big]}}{8\cosh^{2}{\Big[\sqrt{\frac{\Lambda}{3}}t_0\Big]}}\nonumber\\&\hspace{0.25in}+\Big\{\frac{1}{4}-\log{4}\Big\}\tanh^{2}{\Big[\sqrt{\frac{\Lambda}{3}}t_0\Big]}\Bigg\}\\
\tilde{S}&=\frac{4\pi4!}{2\kappa\Lambda}\Bigg\{ \frac{1}{2}\log{\Bigg|\frac{1+e^{-2\sqrt{\frac{\Lambda}{3}}t_{0}}}{2}\Bigg|}+\frac{1-\sinh^{2}{\Big[\sqrt{\frac{\Lambda}{3}}t_0\Big]}}{8\cosh^{2}{\Big[\sqrt{\frac{\Lambda}{3}}t_0\Big]}}\nonumber\\&\hspace{0.25in}+\frac{1}{2}\sqrt{\frac{\Lambda}{3}}t_{0}+\Big\{\frac{1}{4}-\log{4}\Big\}\tanh^{2}{\Big[\sqrt{\frac{\Lambda}{3}}t_0\Big]}\Bigg\}\nonumber
\end{align}


Fourier transforming the amplitude with $\tilde{S} = S - S_{0}$ and employing a similar expansion as (\ref{1p1pole}) reveals the pole again,
\begin{align}\label{3p1pole}
&\int^{\infty}_{0}d\,t_{0}\,e^{i (\tilde{S}[t_0] - \omega\,t_0)}=\int^{\infty}_{0}d\,t_{0}\, e^{i(2\frac{4\pi}{\kappa}\sqrt{\frac{3}{\lambda}}t_{0}-\omega t_0)}\Big(1\nonumber\\&\hspace{0.1in}+i\frac{3\cdot 2}{\Lambda}\frac{4\pi}{\kappa}\Bigg\{\log{\Big|\frac{1+e^{-2\sqrt{\frac{\Lambda}{3}}t_{0}}}{2}\Big|}+\frac{1-\sinh^{2}{\Big[\sqrt{\frac{\Lambda}{3}}t_0\Big]}}{8\cosh^{2}{\Big[\sqrt{\frac{\Lambda}{3}}t_0\Big]}}\nonumber\\&\hspace{0.2in}+\Big\{\frac{1}{4}-\log{4}\Big\}\tanh^{2}{\Big[\sqrt{\frac{\Lambda}{3}}t_0\Big]}\Bigg\}+\ldots\Big)\nonumber\\&\hspace{0.3in}= \frac{i}{\omega -2\frac{4\pi}{\kappa}\sqrt{\frac{3}{\Lambda}}} + \rho_{1}[\omega] +\ldots.
\end{align}

Again we have a pole in the spectral representation at the energy of the static patch, $2\frac{4\pi}{\kappa}\sqrt{\frac{3}{\Lambda}}$. This term is present in  $d+1$ dimensions. The pole in (\ref{3p1pole}) occurs again at a real value of $\omega$ but this is an approximation. This pole is also shifted by a slightly imaginary amount, which is standard in the analysis of resonances \cite{0201503972,2008AIPC.1077...31R}. To the order we are studying here the rate is just that of the standard CDL instanton \cite{Freivogel:2004rd,Freivogel:2006xu,Guth:2007ng}.  

\section{Discussions and Conclusions}
In this paper we presented the technical details of the computations summarized in \cite{Maltz:2016iaw}. The main implication  of this is the following: There exist transition amplitudes between excited states of supersymmetric flat vacua employed in string theory, that possess dS vacua as resonances. Although we have not mentioned it a given dS vacuum contains an exponentially large number of almost degenerate states and in a real quantum theory we would expect a correspondingly dense collection of poles. This is analogous to the idea of a  black hole as a collection of resonances. Deforming the CDL instanton of \cite{Freivogel:2004rd} to a \emph{constrained} CDL instanton solution, allowed us to restrict the path integral over all histories of a transition amplitude between supersymmtric flat vacua to histories were only the time between the nucleation points was integrated over. The spectral representation of this amplitude  possesses a pole indicative of dS resonances for D=2,4. In fact as the pole comes from the linear $t_0$ growth of the action contribution of the bulk volume of the causal patch, it is likely that the pole occurs in $dS_{D}$. The deformation of the original CDL instanton respects an $O(D-2)$ subgroup of the instanton's $O(D-1)$ symmetry as the volume determinate factorizes into a $t_0$ dependent piece and the transverse $\mathbb{S}^{D-2}$; therefore, barring technical issues the same analysis can be carried out in $D$ dimensions such as $D=10,11D$.

None of this should be taken to mean that ordinary scattering amplitudes for finite numbers of particles contain dS \footnote{Exact Minkowski space is static and stable. Minkowski space with a finite number of low energy scattering particles would also be static and stable and would not result in a dS. The transition we are referring to here is between an infinite number of particles in highly fine-tuned states that can be thought of as domain wall to domain wall transitions.}. The $|\ein\rangle$ and $|\eout\rangle$ states we are discussing are open (k=-1) FRW cosmologies that contain an infinite number of particles. The particles are uniformly distributed on hyperbolic surfaces and, in particular, there exists an infinite number of particles on $\Sigma$ of Fig. \ref{CDLtoconstrainedCDL} (left). This suggests that states of this type form a superselection sector in which the dS resonances are found. Since these states contain an infinite number particles but their entropy must not exceed the finite dS entropy of the causal patch, they must be infinitely fine tuned. Such states would be the bulk states of FRW/CFT \cite{Freivogel:2004rd,Freivogel:2006xu,Susskind:2007pv,Sekino:2009kv} or similar string theory construction that possesses dS as an intermediate configuration. One should also point out that the super-selection sector of states of this type may not be continuously connected as in standard S-matrix amplitudes. If an ``off shell" history in the transition amplitude is not in the superselection sector proposed here it is very likely that it will cause a crunch as opposed to a dS \cite{Banks:2002fe,Bousso:2006ge,Bousso:2010zi} or some other unknown configuration that is not a small perturbation of the semiclassical spacetime. In the analysis we employed here, we assume we have restricted to states that do not crunch. The infinitely fine-tuned nature of these states suggests there is a large but finite number of them, essentially the exponential of the dS entropy, $\sim e^{10^{120}}$.  Choosing  \ein and \eout states that do not crunch is just one more criterion for selecting appropriate states that lead to a dS as opposed to another spacetime and more analysis is needed on this point.

It has been asked how recent work on complexity and relations between geometry and entanglement apply in a cosmological setting.
In $2+1$ dimensions the action calculation when continued to AdS is similar to wick rotated calculations relating complexity  to action in the AdS BTZ black hole \cite{Brown:2015bva,Brown:2015lvg}; see Fig. \ref{spacetimeying}. In the continuation $\mathcal{V}$ replaces the Wheeler de Witt patch of \cite{Brown:2015bva,Brown:2015lvg}.  In both cases the action grows linearly with time $t_0$, which in the dS case leads to the resonant pole found; in the AdS version it represents the linear growth in complexity.  It is possible that in cosmology the exponential expansion of space may also represent a growth in complexity. This is analogous to the growth of complexity being related to the lengthening of nontransversable wormhole throats in the AdS BTZ setting. Further study in this direction is demanded.


\begin{acknowledgments}
The author thanks Leonard Susskind and Ying Zhao for extremely helpful discussions in the course of this work. Furthermore the author thanks  Ahmed Almheiri, Dionysios Anninos, Tom Banks, Ning Bao, Adam Brown, David Berenstein, Evan Berkowitz, Raphael Bousso, Ben Frivogel, Ori Ganor, Masanori Hanada, Stefan Leichenauer, Don Page, Micheal Peskin, Douglas Stanford, Raphael Sgier, Yasuhiro Sekino, and Jason Weinberg  for stimulating discussions and comments. The author also thanks Tim Mernard and Nicholas Johnson for stimulating discussions and hospitality during the completion of this work. The work of the author is supported by the California Alliance fellowship (NSF Grant No. 32540).
\end{acknowledgments}


\appendix

\section{Junction Conditions and the \emph{constrained} Geometry}\label{geom}

\hspace{0.1in}In this appendix we  construct the \emph{constrained} CDL instanton geometry. For a given value of $\eta_0$ the \emph{constrained} CDL can be viewed as an ``off shell" path of the path integral (\ref{calschematic}). 
 The \emph{constrained} CDL in the limit of going ``on shell" ($\eta_0\rightarrow 0$) becomes the CDL instanton with a zero tension domain wall; ``off shell" the domain walls are null; see Fig. \ref{constrainedCDL}. With the gauge choice that the separated bubbles are centered on the de Sitter coordinate $\psi =0$ with the future bubble nucleation at $\eta_0$ and the past bubble ending at the time reversed $-\eta_0$. 

To do this within the context of GR we will employ the Barrab\'{e}s-Israel null junction conditions \cite{PhysRevD.43.1129}\footnote{Reviews on the null junction conditions and how to deal with null boundaries in GR are \cite{PhysRevD.43.1129,Parattu:2015gga,Parattu:2016trq,poisson,1966NCimB..44....1I,Musgrave:1995ka,0264-9381-14-5-029}.}.
The de Sitter metric in conformal coordinates is
\begin{equation}\label{dsMetconf}
ds^2_{dS} = \frac{3}{\Lambda \cos^{2}{\eta}}\Big\{-d\eta^2 + d\psi^2 + \sin^{2}{\psi}\dOmt\Big\},
\end{equation}

where $\Lambda$ is the cosmological constant related to $l_{dS}$ by $l_{dS} =\sqrt{\frac{3}{\Lambda}}$. 
The coordinates $-\frac{\pi}{2}\leq \eta\leq\frac{\pi}{2}$ and $0\leq \psi \leq \pi$ along with the sphere's coordinates cover the entire de Sitter spacetime.  

An open hyperbolic FRW universe with $\Lambda = 0$ ``\emph{hat}" with no matter has the metric

\begin{equation}\label{FRWtchi}
ds^{2}_{\text{FRW-Milne}} = -d\tau^2 +\tau^{2}\Big(d\chi^2 + \sinh^{2}{\chi}\dOmt\Big),
\end{equation}

with $0\leq\tau<\infty$ and $0\leq\chi<\infty$.

This spacetime is also known as the Milne universe \cite{milne1935relativity}. It is just the interior of the forward light cone of the origin in Minkowski space, as can be seen via the coordinate change $t = \tau\cosh{\chi}$, $r = \tau\sinh{\chi}$ resulting in

\begin{equation}\label{minkowski}
ds^2 = -dt^2 +dr^2 +r^{2}\dOmt
\end{equation}

with $r\leq t$. For later convenience we perform the change of variables $r = \sqrt{\frac{3}{\Lambda}}\frac{\sin{\psi}}{\cos{[\psi+\eta_0]}}$, which results in the $\mathbb{S}^2$ of both the hat and the de Sitter spacetimes having the same radial coordinate,

\begin{equation}\label{hat}
ds^{2}_{hat} = -dt^2 + \frac{3\cos^{2}{\eta_0}}{\Lambda\cos^{4}{[\psi +\eta_0]}}d\psi^2 +\frac{3\sin^{2}{\psi}}{\Lambda\cos^{4}{[\psi+\eta_0]}}\dOmt.\nonumber
\end{equation}

We employ this form of the metric while stitching to de Sitter. In these coordinates there is not a coordinate singularity along the stitching surface $t=r=\sqrt{\frac{3}{\Lambda}}\frac{\sin{\psi}}{\cos^{2}{(\psi+\eta_0)}}$, which in (\ref{FRWtchi}) is the line coordinate singularity $\tau = 0$. 
 
For a nice review on how to use the junction conditions to stitch together spacetime on null surfaces the reader is encouraged to look at \cite{PhysRevD.43.1129,poisson}.
The future FRW hat, which we refer to as region \RI\,, 
to keep in line with the notation of \cite{Freivogel:2006xu}\footnote{In \cite{Freivogel:2004rd} the labeling of the regions is shuffled with $III \rightarrow I$, $I\rightarrow II$, and $II\rightarrow III$. To avoid confusion we follow the conventions of \cite{Freivogel:2006xu}.}, is connected to the dS on the null line, $t-r=0$ in the hat, and $0= -\psi +\eta -\eta_0$ in the dS. This is referred to as the future null boundary (\textbf{F.B.}), see Fig. \ref{constrainedCDL}.

Following the junction conditions \cite{PhysRevD.43.1129},
we decompose the metric into
\begin{align}
g_{\mu\nu} &= - \tilde{\eta}(n_{\mu}N_{\nu}+n_{\nu}N_{\mu}) + e^{A}_{\mu}e^{B}_{\nu}\sigma_{AB}\nonumber\\ &= - \tilde{\eta}(n_{\mu}N_{\nu}+n_{\nu}N_{\mu}) + e^{a}_{\mu}e^{b}_{\nu}h_{ab}
\end{align}
with the null normal (surface gradient)  $n_{\mu} = \alpha^{-1}\partial_{\mu} \Phi$ and null auxiliary vector $N^{\mu}$. $\tilde{\eta}^{-1}$ is not the coordinate $\eta$ but a real constant. In order to form a complete basis  for the metric we must also enforce the condition that $n\cdot N = \tilde{\eta}^{-1}$ across the boundary as well as $n\cdot e^{A} =0$ and $N\cdot e^{A}=0$ on the boundary. $\Phi[x]$ is a scalar function of the coordinates and $\Phi[x^{\mu}] =0$ defines the null surface that we are joining the metrics along. The projection of the auxiliary vector to the surface $N_{a} =N_{\mu}e^{\mu}_{a}$ must be continuous across the boundary. Enforcing $n\cdot N = \tilde{\eta}^{-1}$ across the boundary determines $\alpha$ \footnote{$\alpha$ is the relative scale between the auxiliary vector $N^{\mu}$ on opposite sides of the surface. Once a value of $\alpha$ is chosen on one side, the junction condition $N_{\mu}e^{\mu}_{a}|_{+}-N_{\mu}e^{\mu}_{a}|_{-} = 0$, determines $\alpha$ on the other side \cite{PhysRevD.43.1129}. Since both $n^{\mu}$ and $N^{\mu}$ are null the initial choice of $\alpha$ is arbitrary and by convention is negative for nontimelike surfaces. Here we use the conventions of \cite{PhysRevD.43.1129}}. $n_{\mu}=\alpha^{-1}\partial_{\mu}\Phi$ with $\alpha=-1$ in region \RIII\, (dS region) results in $\alpha =-\sqrt{\frac{3}{\Lambda}}\frac{1}{\cos{\eta_0}}$ in regions \RI\, and \RII.   For the F.B., we have $\Phi_{+} =t- \sqrt{\frac{3}{\Lambda}}\frac{\sin{\psi}}{\cos^{2}{[\psi+\eta_0]}}$ in the ``hat" coordinates and $\Phi_{+} = -\psi +(\eta - \eta_0)$ in the dS coordinates. The null auxiliary vector is defined by $N^{\mu}n_{\mu}= \tilde{\eta}^{-1} =-1$.
\begin{center}
\begin{align}\label{nullauxvectorsfuture}
&\textbf{\RI:}\\&\hspace{0.1in}n^{\mu}\partial_{\mu}=-\sqrt{\frac{\Lambda}{3}}\cos{\eta_0}\Bigg(\partial_{t}-\sqrt{\frac{\Lambda}{3}}\frac{\cos^{2}{[\psi+\eta_0]}}{\cos{\eta_0}}\partial_{\psi}\Bigg)\nonumber\\&\hspace{0.1in}n_{\mu}dx^{\mu}=-\sqrt{\frac{\Lambda}{3}}\cos{\eta_0}\Bigg(-dt-\sqrt{\frac{\Lambda}{3}}\frac{\cos^{2}{[\psi+\eta_0]}}{\cos{\eta_0}}\Bigg)\nonumber\\&\hspace{0.125in}N^{\mu}\partial_{\mu} = -\sqrt{\frac{3}{\Lambda}}\frac{1}{\cos{\eta_0}}\Bigg(\frac{1}{2}\partial_{\tau} +\frac{1}{2}\sqrt{\frac{\Lambda}{3}}\frac{cos^{2}{[\psi+\eta_0]}}{cos{\eta_0}}\partial_{\psi}\Bigg)\nonumber\\&\hspace{0.1in}N_{\mu}dx^{\mu}=-\sqrt{\frac{3}{\Lambda}}\frac{1}{\cos{\eta_0}}\Bigg(\frac{1}{2}dt+\frac{1}{2}\sqrt{\frac{3}{\Lambda}}\frac{\cos{\eta_0}}{\cos^{2}{[\psi +\eta_0]}}\Bigg)\nonumber
\end{align}
\begin{align}
&\textbf{F.B. of region \RIII:}\\&\hspace{0.1in}n^{\mu}\partial_{\mu} = \frac{\Lambda}{3}\cos^{2}{[\psi+\eta_0]}\partial_\eta +\frac{\Lambda}{3}\cos^{2}{[\psi+\eta_0]}\partial_\psi\nonumber\\&\hspace{0.1in}n_{\mu}dx^{\mu} = -d\eta +d\psi\nonumber\\&\hspace{0.1in}N^{\mu}\partial_{\mu} =\frac{1}{2}\partial_{\eta}-\frac{1}{2}\partial_{\psi}\nonumber\\&\hspace{0.1in}N_{\mu}dx^{\mu} = \frac{-3\,d\eta}{2\Lambda\cos^{2}{[\psi+\eta_0]}} +\frac{-3\,d\psi}{2\Lambda\cos^{2}{[\psi+\eta_0]}}\nonumber.
\end{align}
\end{center}
We  employ $\xi^{a} =(\psi,\theta,\phi)$ as the intrinsic coordinates on the null surface and express $n^{\mu}$ in the basis of null generators \cite{PhysRevD.43.1129} 
$n^{\mu} = l^{a}e^{\mu}_{a}$ as follows

\begin{equation}\label{projectionvectorsfuture}
\textbf{F.B.}\hspace{0.45in}l^{a}=\Bigg(\frac{\Lambda}{3}\cos^{2}{[\psi+\eta_0]},0,0\Bigg),
\end{equation}
with $e^{\mu} = \frac{\partial x^{\mu}}{\partial\xi^{a}}$ and $x^{\mu}$ being the coordinates  of the spacetime regions on either side of the boundary.

This choice of $l^{a}$ allows us to define $h^{ab}_{*}$, which satisfies the following relation with the surface's degenerate three metric \cite{PhysRevD.43.1129}, $h_{ab}$, 
\begin{equation}\label{degeneratemetric}
h^{ac}_{*}h_{bc} = \delta^{a}_{b} + \tilde{\eta}l^{a}N_{\mu}e^{\mu}_{b},
\end{equation}

resulting in the degenerate three metric $h_{ab}$ and $h^{ab}_{*}$ being of the block diagonal form 
\begin{equation}\label{block}
h_{ab} = \begin{bmatrix}
    0& 0 \\
    0&\sigma_{AB}
\end{bmatrix}\hspace{0.25in}
h^{ab}_{*} = \begin{bmatrix}
    0& 0 \\
    0&\sigma^{AB}
\end{bmatrix}
\end{equation}
with $\sigma_{AB}$,  $A,B = (\theta,\phi)$ being the metric of $\mathbb{S}^2$ with radius $r=\sqrt{\frac{3}{\Lambda}}\frac{\sin{\psi}}{\cos^{2}{(\psi + \eta_0)}}$,
\begin{equation}\label{2sphere}
ds^{2} = \sigma_{AB}d\theta^{A}d\theta^{B} = \frac{3\sin^{2}{\psi}}{\Lambda\cos^{4}{(\psi+\eta_0)}}\dOmt,
\end{equation}
yielding $h_{ab} = e^{A}_{a}e^{B}_{b}\sigma_{AB}$,  $h^{ab}_{*} =e^{a}_{A}e^{b}_{B}\sigma^{AB}$ \footnote{Since $h_{ab}$ is the metric of a null surface it is degenerate ($\det{h} = 0$) and therefore does not possess an inverse metric $h^{ab}$. (\ref{degeneratemetric}) only determines $h^{ab}_{*}$ up to a gauge $h^{ab}_{*} \rightarrow h^{ab}_{*} + 2\lambda l^{a}l^{b}$, with $\lambda$ being an arbitrary function \cite{PhysRevD.43.1129}. Here we make the gauge choice that both $h_{ab}$ and $h^{ab}_{*}$ have the block diagonal form (\ref{block}).}.

Similarly the past hat is stitched onto the surface $\Phi_{-} = t+r = 0$ in the hat coordinates $\Phi_{-}=\psi +(\eta+\eta_0) = 0$ in the dS coordinates \footnote{While the metric for the past hat has the same form as (\ref{hat}) it should be noted that $t$ now has the coordinate range $-\infty<t\leq0$.}. For completeness we  give the $n^{\mu}$, $N^{\mu}$, and $l^{a}$ for the past hat.

\begin{center}
\begin{align}\label{nullauxvectorsfuture2}
&\textbf{P.B. of region \RIII:}\\&\hspace{0.1in}n^{\mu}\partial_{\mu} = \frac{\Lambda}{3}\cos^{2}{[\psi+\eta_0]}\partial_\eta -\frac{\Lambda}{3}\cos^{2}{[\psi+\eta_0]}\partial_\psi\nonumber\\&\hspace{0.1in}n_{\mu}dx^{\mu} = -d\eta -d\psi\nonumber\\&\hspace{0.1in}N^{\mu}\partial_{\mu} =\frac{1}{2}\partial_{\eta}+\frac{1}{2}\partial_{\psi}\nonumber\\&\hspace{0.1in}N_{\mu}dx^{\mu} = \frac{-3\,d\eta}{2\Lambda\cos^{2}{[\psi+\eta_0]}} +\frac{3\,d\psi}{2\Lambda\cos^{2}{[\psi+\eta_0]}}\nonumber\\
&\textbf{\RII:}\nonumber\\&\hspace{0.1in}n^{\mu}\partial_{\mu}=-\sqrt{\frac{\Lambda}{3}}\cos{\eta_0}\Bigg(-\partial_{t}+\sqrt{\frac{\Lambda}{3}}\frac{\cos^{2}{[\psi+\eta_0]}}{\cos{\eta_0}}\partial_{\psi}\Bigg)\nonumber\\&\hspace{0.1in}n_{\mu}dx^{\mu}=-\sqrt{\frac{\Lambda}{3}}\cos{\eta_0}\Bigg(dt+\sqrt{\frac{\Lambda}{3}}\frac{\cos^{2}{[\psi+\eta_0]}}{\cos{\eta_0}}\Bigg)\nonumber\\&\hspace{0.125in}N^{\mu}\partial_{\mu} = \sqrt{\frac{3}{\Lambda}}\frac{1}{\cos{\eta_0}}\Bigg(\frac{1}{2}\partial_{\tau}+\frac{1}{2}\sqrt{\frac{\Lambda}{3}}\frac{cos^{2}{[\psi+\eta_0]}}{cos{\eta_0}}\partial_{\psi}\Bigg)\nonumber
\end{align}
\begin{align}
&\hspace{0.1in}N_{\mu}dx^{\mu}=-\sqrt{\frac{3}{\Lambda}}\frac{1}{\cos{\eta_0}}\Bigg(\frac{1}{2}dt-\frac{1}{2}\sqrt{\frac{3}{\Lambda}}\frac{\cos{\eta_0}}{\cos^{2}{[\psi +\eta_0]}}\Bigg).\nonumber
\end{align}
\end{center}

\begin{equation}\label{projectionvectorsfuture2}
\textbf{P.B.}\hspace{0.45in}l^{a}=\Big(-\frac{\Lambda}{3}\cos^{2}{[\psi+\eta_0]},0,0\Big).
\end{equation}

Since these are null shells, the junction conditions require us to compute the discontinuity in the \emph{transverse} extrinsic curvature $\mathpzc{K}_{ab} = -N_{\mu}e^{\nu}_{b}\nabla_{\nu}e^{\mu}_{a} = \mathpzc{K}_{ba}$ to determine the stress tensor required to support this geometry\footnote{Since the surface is null, its normal $n_{\mu}$ is  orthogonal to the surface yet the resulting vector $n^{\mu}$ is parallel to the surface, due to $n^{\mu}n_{\mu}=0$. This means that the standard extrinsic curvature $K_{ab} = -n_{\mu}e^{\nu}_{b}\nabla_{\nu}e^{\mu}_{a}$ does not carry any transverse information on a null surface. This is why the transverse extrinsic curvature must be introduced for the stitching \cite{PhysRevD.43.1129}.}, defining the symbol $\gamma_{ab} = \mathpzc{K}_{ab}|_{+}-\mathpzc{K}_{ab}|_{-}$ to  be the difference of $\mathpzc{K}_{ab}$ on both sides of the stitching surface evaluated at the surface, in their respective coordinate charts. We can define the surface stress tensor $S^{ab}$, which has the following relation on null shells \cite{PhysRevD.43.1129,PhysRevD.39.2901}, 
\begin{align}\label{surfacestress}
-16\pi S^{ab} = \Big(&g^{ac}_{*}l^{b}l^{d}+g^{bd}_{*}l^{a}l^{c}\nonumber\\&\hspace{0.1in}-g^{ab}_{*}l^{c}l^{d}-g^{cd}_{*}l^{a}l^{b}\Big)\gamma_{cd}.
\end{align}

Employing (\ref{nullauxvectorsfuture}-\ref{surfacestress}) we have

\begin{equation}
S^{ab} = -\frac{1}{8\pi}\Bigg(\frac{\sin{\psi}}{\cos{\eta_0}\cos{(\psi +\eta_0)}}\Bigg)l^{a}l^{b}.
\end{equation}

The full stress tensor is $T^{\mu\nu} =\alpha e^{\mu}_{a} e^{\nu}_{b}S^{ab}\delta(\Phi)$ in each region, which has different representations in each region dependent on the coordinates employed there. We state the stress tensor here in all regions for clarity.

\bwt 
\begin{align}\label{stresstensor}
&\textbf{\RI}\nonumber\\\hspace{0.2in}&T^{\mu\nu}\partial_{\mu}\otimes\partial_{\nu}=\frac{1}{8\pi}\Bigg(\sqrt{\frac{\Lambda}{3}}\frac{\sin{\psi}}{cos{[\psi+\eta_0]}}\partial_{\tau}\otimes\partial_{\tau}+\frac{\Lambda\cos{[\psi+\eta_0]}\sin{\psi}}{3\cos{\eta_0}}\Big(\partial_{\tau}\otimes\partial_{\psi}+\partial_{\psi}\otimes\partial_{\tau}\Big)\nonumber\\&\hspace{1.4in}+\Bigg(\frac{\Lambda}{3}\Bigg)^{3/2}\frac{\cos^{3}{[\psi+\eta_0]}\sin{\psi}}{\cos^{3}{\eta_0}}\partial_{\psi}\otimes\partial_{\psi}\Bigg)\delta\Bigg[\tau-\sqrt{\frac{3}{\Lambda}}\frac{\sin{\psi}}{\cos{[\psi+\eta_0]}}\Bigg]\\
&\textbf{\RIII}\nonumber\\\hspace{0.2in}&T^{\mu\nu}\partial_{\mu}\otimes\partial_{\nu} = \frac{1}{8\pi}\Bigg(\frac{\Lambda}{3}\Bigg)^{2}\frac{cos^{3}{[\psi +\eta_0]}\sin{\psi}}{\cos{\eta_0}}\Big(\partial_{\eta}\otimes\partial_{\eta}+\partial_{\eta}\otimes\partial_{\psi}+\partial_{\psi}\otimes\partial_{\eta}+\partial_{\psi}\otimes\partial_{\psi}\Big)\nonumber\\&\hspace{1.1in}\times\delta(\eta -\eta_0 -\psi)+\frac{1}{8\pi}\Bigg(\frac{\Lambda}{3}\Bigg)^{2}\frac{cos^{3}{[\psi +\eta_0]}\sin{\psi}}{\cos{\eta_0}}\Big(\partial_{\eta}\otimes\partial_{\eta}\nonumber\\&\hspace{1.3in}-\partial_{\eta}\otimes\partial_{\psi}-\partial_{\psi}\otimes\partial_{\eta}+\partial_{\psi}\otimes\partial_{\psi}\Big)\delta[\eta +\eta_0 +\psi] - \frac{\Lambda}{8\pi}g^{\mu\nu}_{dS}\partial_{\mu}\otimes\partial_{\nu}\label{a}\\
&\textbf{\RII}\nonumber\\\hspace{0.2in}&T^{\mu\nu}\partial_{\mu}\otimes\partial_{\nu}=\frac{1}{8\pi}\Bigg(\sqrt{\frac{\Lambda}{3}}\frac{\sin{\psi}}{cos{[\psi+\eta_0]}}\partial_{\tau}\otimes\partial_{\tau}-\frac{\Lambda\cos{[\psi+\eta_0]}\sin{\psi}}{3\cos{\eta_0}}\Big(\partial_{\tau}\otimes\partial_{\psi}+\partial_{\psi}\otimes\partial_{\tau}\Big)\nonumber\\&\hspace{1.4in}+\Bigg(\frac{\Lambda}{3}\Bigg)^{3/2}\frac{\cos^{3}{[\psi+\eta_0]}\sin{\psi}}{\cos^{3}{\eta_0}}\partial_{\psi}\otimes\partial_{\psi}\Bigg)\delta\Bigg[\tau+\sqrt{\frac{3}{\Lambda}}\frac{\sin{\psi}}{\cos{[\psi+\eta_0]}}\Bigg]\label{b}
\end{align}.
\ewt

As was argued in the main text, while the stress tensor in this coordinate representation does depend on the time $\eta_0$, the boost invariance of the geometry implies that the action contribution from the stress tensor should not depend on the time $\eta_0$.

We can now define the ``off-shell" Coleman-De Luccia geometry as
\begin{align}\label{metricdist}
g_{\mu\nu} &= g^{\text{(\RI)}}_{\mu,\nu}\Theta[\Phi_{+}]+ g^{\text{(\RIII})}_{\mu,\nu}\Theta[-\Phi_{+}]\Theta[\Phi_{-}]\nonumber\\&\hspace{0.5in} + g^{\text{(\RII)}}_{\mu,\nu}\Theta[-\Phi_{-}].
\end{align} Here $g_{\mu\nu}$ is expressed as a distribution employing $\Theta[x]$, which is the Heaviside theta function with $\Theta[x]= 1$ for $x>0$,$\Theta[x]=0$ for $x<0$ and $\Theta[0]=1/2$. (\ref{metricdist}) along with the 
stress tensor (\ref{stresstensor})-(\ref{b}) represents the spacetime.

\section{Justification For the Integration Region}\label{stichjust}

\hspace{0.1in}In this section, we argue that in the approximations we have made the integration region used is the only one necessary to calculate the action  of the causal patch. To begin assume that we have a global coordinate chart for the entire \emph{constrained} CDL spacetime. Such a chart exists because the stitched spacetime foliated by $\mathbb{S}^2$s is topologically simple. One can construct such a chart system by using skew-Gaussian coordinates attached to geodesics that reach into all three regions and are maximally smooth \cite{PhysRevD.43.1129}.  The metric for this entire spacetime can then be written as a Dirac distribution treating the domain walls as thin shells

\begin{equation}\label{diracglobmet}
g_{\alpha\beta}=g^{(1)}_{\alpha\beta}\,\Theta[\Phi_{1}]+g^{(2)}_{\alpha\beta}\,\Theta[-\Phi_1]\Theta[\Phi_2]+g^{(3)}_{\alpha\beta}\,\Theta[-\Phi_2].
\end{equation}

Here $\Theta[x]$ is the Heaviside theta function with $\Theta[x]= 1$ for $x>0$,$\Theta[x]=0$ for $x<0$ and $\Theta[0]=1/2$. The superscripts $1,2,3$ refer to the future Hat, de Sitter, and past Hat regions, respectfully \footnote{Note that this index is not the same as the convention we have been using throughout the paper and is employed because the dS region includes regions \RIII - \RV.} and $\Phi_{1(2)}$ are the scalar equations that vanish on the domain walls of the future and past hat regions, respectfully. They are the analog of $\Phi_+= \eta -(\psi+\eta_0)$ and $\Phi_- = \eta + (\psi+\eta_0)$  that were employed in the main text.

Following the formulation of the junction conditions in \cite{PhysRevD.43.1129} we can construct the distributions for Christoffel symbols,

\bwt
\begin{align}
2\Gamma_{\sigma\alpha\beta}&=\partial_{\alpha}g_{\beta\sigma}+\partial_{\beta}g_{\sigma\alpha}-\partial_{\sigma}g_{\alpha\beta}\nonumber\\&=\Big(\partial_{\alpha}g^{(1)}_{\beta\sigma}+\partial_{\beta}g^{(1)}_{\sigma\alpha}-\partial_{\sigma}g^{(1)}_{\alpha\beta}\Big)\Theta[\Phi_1]+\Big(\partial_{\alpha}g^{(2)}_{\beta\sigma}+\partial_{\beta}g^{(2)}_{\sigma\alpha}-\partial_{\sigma}g^{(2)}_{\alpha\beta}\Big)\Theta[-\Phi_1]\Theta[\Phi_2]\nonumber\\&\hspace{0.2in}+\Big(\partial_{\alpha}g^{(3)}_{\beta\sigma}+\partial_{\beta}g^{(3)}_{\sigma\alpha}-\partial_{\sigma}g^{(3)}_{\alpha\beta}\Big)\Theta[-\Phi_2]+g^{(1)}_{\alpha\beta}\partial_{\alpha}\Theta[\Phi_1]+g^{(1)}_{\alpha\beta}\partial_{\alpha}\Theta[\Phi_1]\nonumber\\&\hspace{0.2in}-g^{(1)}_{\alpha\beta}\partial_{\alpha}\Theta[\Phi_1]+g^{(2)}_{\alpha\beta}\partial_{\alpha}\big(\Theta[-\Phi_1]\Theta[\Phi_2]\big)+g^{(2)}_{\alpha\beta}\partial_{\alpha}\big(\Theta[-\Phi_1]\Theta[\Phi_2]\big)\nonumber\\&\hspace{0.2in}-g^{(2)}_{\alpha\beta}\partial_{\alpha}\big(\Theta[-\Phi_1]\Theta[\Phi_2]\big)+g^{(2)}_{\alpha\beta}\partial_{\alpha}\Theta[-\Phi_2]+g^{(2)}_{\alpha\beta}\partial_{\alpha}\Theta[-\Phi_2]-g^{(3)}_{\alpha\beta}\partial_{\alpha}\Theta[-\Phi_2]\nonumber
\end{align}
\begin{align}\label{christofelldis}
&=2\Gamma^{(1)}_{\sigma\alpha\beta}\Theta[\Phi_1]+2\Gamma^{(2)}_{\sigma\alpha\beta}\Theta[-\Phi_1]\Theta[\Phi_2]+2\Gamma^{(3)}_{\sigma\alpha\beta}\Theta[-\Phi_2]\nonumber\\&\hspace{0.2in}+g^{(1)}_{\beta\sigma}\delta[\Phi_1]\partial_{\alpha}\Phi_1-g^{(2)}_{\beta\sigma}\delta[-\Phi_1]\Theta[\Phi_2]\partial_{\alpha}\Phi_1 +g^{(2)}_{\beta\sigma}\Theta[-\Phi_1]\delta[\Phi_2]\partial_{\alpha}\Phi_{2}-g^{(3)}_{\beta\sigma}\Theta[-\Phi_2]\partial_{\alpha}\Phi_{2}\nonumber\\&\hspace{0.2in}+g^{(1)}_{\sigma\alpha}\delta[\Phi_1]\partial_{\beta}\Phi_1-g^{(2)}_{\sigma\alpha}\delta[-\Phi_1]\Theta[\Phi_2]\partial_{\beta}\Phi_1 +g^{(2)}_{\sigma\alpha}\Theta[-\Phi_1]\delta[\Phi_2]\partial_{\beta}\Phi_{2}-g^{(3)}_{\sigma\alpha}\Theta[-\Phi_2]\partial_{\beta}\Phi_{2}\nonumber\\&\hspace{0.2in}+g^{(1)}_{\alpha\beta}\delta[\Phi_1]\partial_{\sigma}\Phi_1-g^{(2)}_{\alpha\beta}\delta[-\Phi_1]\Theta[\Phi_2]\partial_{\sigma}\Phi_1 +g^{(2)}_{\alpha\beta}\Theta[-\Phi_1]\delta[\Phi_2]\partial_{\sigma}\Phi_{2}-g^{(3)}_{\alpha\beta}\Theta[-\Phi_2]\partial_{\sigma}\Phi_{2}.\nonumber\\&\,
\end{align}
\ewt
Because of the time ordering $\Theta[\Phi_2]=1$ when $\Phi_1=0$ (the past hat boundary is in the past of the future hat boundary), terms of the form $g^{(1)}_{\beta\sigma}\delta[\Phi_1]\partial_{\alpha}\Phi_1-g^{(2)}_{\beta\sigma}\delta[-\Phi_1]\Theta[\Phi_2]\partial_{\alpha}\Phi_1 = \delta[\Phi_1]\partial_{\alpha}\Phi_1\Big(g^{(1)}_{\beta\sigma}-g^{(2)}_{\beta\sigma}\Big)$. This allows us to rewrite (\ref{christofelldis}) as
\begin{align}\label{christofelldis2}
&2\Gamma_{\sigma\alpha\beta}\nonumber\\&=2\Gamma^{(1)}_{\sigma\alpha\beta}\Theta[\Phi_1]+2\Gamma^{(2)}_{\sigma\alpha\beta}\Theta[-\Phi_1]\Theta[\Phi_2]+2\Gamma^{(3)}_{\sigma\alpha\beta}\Theta[-\Phi_2]\nonumber\\&\,+\delta[\Phi_1]\partial_{\alpha}\Phi_1\Big(g^{(1)}_{\beta\sigma}-g^{(2)}_{\beta\sigma}\Big)+\delta[\Phi_2]\partial_{\alpha}\Phi_2\Big(g^{(2)}_{\beta\sigma}-g^{(3)}_{\beta\sigma}\Big)\nonumber\\&\, +\delta[\Phi_1]\partial_{\beta}\Phi_{1}\Big(g^{(1)}_{\sigma\alpha}-g^{(2)}_{\sigma\alpha}\Big)+\delta[\Phi_2]\partial_{\beta}\Phi_{2}\Big(g^{(2)}_{\sigma\alpha}-g^{(3)}_{\sigma\alpha}\Big)\nonumber\\&\,-\delta[\Phi_1]\partial_{\sigma}\Phi_1\Big(g^{(1)}_{\alpha\beta}-g^{(2)}_{\alpha\beta}\Big)-\delta[\Phi_2]\partial_{\sigma}\Phi_2\Big(g^{(2)}_{\alpha\beta}-g^{(3)}_{\alpha\beta}\Big).\nonumber\\
\end{align}
In this coordinate system terms of the form $\Big(g^{(1)}_{\alpha\beta}-g^{(2)}_{\alpha\beta}\Big)|_{\Phi_1}=0$, since we have made a global coordinate chart that covers the entire spacetime \footnote{In the generic case when employing the junction conditions as in Appendix \ref{geom}, these relations reduce to the statement that the intrinsic metric is the same on both sides of the boundary, $\Big(e^{\alpha}_{A}e^{\beta}_{B}g^{(1)}_{\alpha\beta}-e^{\alpha}_{A}e^{\beta}_{B}g^{(2)}_{\alpha\beta}\Big)|_{\Phi_1}=0$. This  is the first junction condition when coordinate charts are different on opposite sides of the boundary, the mismatch being pure gauge \cite{PhysRevD.43.1129}.}.
(\ref{christofelldis2}) is then reduced to
\begin{align}\label{christofelldis3}
\Gamma_{\sigma\alpha\beta}&=\Gamma^{(1)}_{\sigma\alpha\beta}\Theta[\Phi_1]+\Gamma^{(2)}_{\sigma\alpha\beta}\Theta[-\Phi_1]\Theta[\Phi_2]+\Gamma^{(3)}_{\sigma\alpha\beta}\Theta[-\Phi_2]\nonumber\\\,\Gamma^{\rho}_{\alpha\beta}&=g^{\rho\sigma}\Gamma_{\sigma\alpha\beta}\nonumber\\&=\Gamma^{(1)\rho}_{\alpha\beta}\Theta[\Phi_1]+\Gamma^{(2)\rho}_{\alpha\beta}\Theta[-\Phi_1]\Theta[\Phi_2]+\Gamma^{(3)\rho}_{\alpha\beta}\Theta[-\Phi_2],
\end{align}
where in the second line we have used the identities $(\Theta[x])^{2}=\Theta[x]$ for $x\neq 0$ and $\Theta[x]\Theta[-x]=0$ for $x \neq 0$.

The Ricci tensor is defined as
\begin{equation}\label{RicciTens}
R_{\mu\nu}=R^{\rho}_{\,\,\,\mu\rho\nu}=\partial_{\rho}\Gamma^{\rho}_{\nu\mu}-\partial_{\nu}\Gamma^{\rho}_{\rho\mu} +\Gamma^{\rho}_{\rho\lambda}\Gamma^{\lambda}_{\nu\mu}-\Gamma^{\rho}_{\nu\lambda}\Gamma^{\lambda}_{\rho\mu},
\end{equation}

which with (\ref{christofelldis3}) yields the following Dirac distribution,

\bwt
\begin{align}\label{riccipieces}
R_{\mu\nu} &=\Big(\partial_{\rho}\Gamma^{(1)\rho}_{\nu\mu}-\partial_{\nu}\Gamma^{(1)\rho}_{\rho\mu}\Big)\Theta[\Phi_1]+\Big(\partial_{\rho}\Gamma^{(2)\rho}_{\nu\mu}-\partial_{\nu}\Gamma^{(2)\rho}_{\rho\mu}\Big)\Theta[-\Phi_1]\Theta[\Phi_2]\nonumber\\&\,+\Big(\partial_{\rho}\Gamma^{(3)\rho}_{\nu\mu}-\partial_{\nu}\Gamma^{(3)\rho}_{\rho\mu}\Big)\Theta[-\Phi_2]+\Big(\Gamma^{(1)\rho}_{\rho\lambda}\Gamma^{(1)\lambda}_{\nu\mu}-\Gamma^{(1)\rho}_{\nu\lambda}\Gamma^{(1)\lambda}_{\rho\mu}\Big)\Theta[\Phi_1]\nonumber\\&\,+\Big(\Gamma^{(2)\rho}_{\rho\lambda}\Gamma^{(2)\lambda}_{\nu\mu}-\Gamma^{(2)\rho}_{\nu\lambda}\Gamma^{(2)\lambda}_{\rho\mu}\Big)\Theta[-\Phi_1]\Theta[\Phi_2]+\Big(\Gamma^{(3)\rho}_{\rho\lambda}\Gamma^{(3)\lambda}_{\nu\mu}-\Gamma^{(3)\rho}_{\nu\lambda}\Gamma^{(3)\lambda}_{\rho\mu}\Big)\Theta[-\Phi_2]\nonumber\\&\,+\Big(\Gamma^{(1)\rho}_{\nu\mu}\delta[\Phi_1]\partial_{\rho}\Phi_1 - \Gamma^{(1)\rho}_{\rho\mu}\delta[\Phi_1]\partial_{\nu}\Phi_1\Big)-\Big(\Gamma^{(2)\rho}_{\nu\mu}\delta[-\Phi_1]\Theta[\Phi_2]\partial_{\rho}\Phi_1\nonumber\\&\, - \Gamma^{(2)\rho}_{\rho\mu}\delta[-\Phi_1]\Theta[\Phi_2]\partial_{\nu}\Phi_1\Big)+\Big(\Gamma^{(2)\rho}_{\nu\mu}\delta[\Phi_2]\Theta[-\Phi_1]\partial_{\rho}\Phi_2 - \Gamma^{(2)\rho}_{\rho\mu}\delta[\Phi_2]\Theta[-\Phi_1]\partial_{\nu}\Phi_2\Big)\nonumber\\&\,-\Big(\Gamma^{(3)\rho}_{\nu\mu}\delta[-\Phi_2]\partial_{\rho}\Phi_2 - \Gamma^{(3)\rho}_{\rho\mu}\delta[-\Phi_2]\partial_{\nu}\Phi_2\Big).
\end{align}
\ewt

Using the definition (\ref{RicciTens}) we can combine the terms in (\ref{riccipieces}) to yield
\bwt
\begin{align}\label{riccitcal}
&R_{\mu\nu} = R^{(1)}_{\mu\nu}\Theta[\Phi_1] + R^{(2)}_{\mu\nu}\Theta[-\Phi_1]\Theta[\Phi_2]+R^{(3)}_{\mu\nu}\Theta[-\Phi_2]+\Big\{\Big(\Gamma^{(1)\rho}_{\nu\mu}-\Gamma^{(2)\rho}_{\nu\mu}\Big)\partial_{\rho}\Phi_{1}-\Big(\Gamma^{(1)\rho}_{\rho\mu}-\Gamma^{(2)\rho}_{\rho\mu}\Big)\partial_{\nu}\Phi_{1}\Big\}\delta[\Phi_1]\nonumber\\&\hspace{2in}+\Big\{\Big(\Gamma^{(2)\rho}_{\nu\mu}-\Gamma^{(3)\rho}_{\nu\mu}\Big)\partial_{\rho}\Phi_{2}-\Big(\Gamma^{(2)\rho}_{\rho\mu}-\Gamma^{(3)\rho}_{\rho\mu}\Big)\partial_{\nu}\Phi_{2}\Big\}\delta[\Phi_2],
\end{align}
\ewt
\bwt
\begin{align}\label{ricciscal}
R = g^{\mu\nu}R_{\mu\nu}&=R^{(1)}\Theta[\Phi_1]+R^{(2)}\Theta[-\Phi_1]\Theta[\Phi_2]+R^{(3)}\Theta[-\Phi_2]\nonumber\\&\,\hspace{0.25in}+\Big\{g^{(1)\mu\nu}\Theta[\Phi_1]+g^{(2)\mu\nu}\Theta[-\Phi_1]\Theta[\Phi_2]\Big\}\Big\{\Big(\Gamma^{(1)\rho}_{\nu\mu}-\Gamma^{(2)\rho}_{\nu\mu}\Big)\partial_{\rho}\Phi_{1}-\Big(\Gamma^{(1)\rho}_{\rho\mu}-\Gamma^{(2)\rho}_{\rho\mu}\Big)\partial_{\nu}\Phi_{1}\Big\}\delta[\Phi_1]\nonumber\\&\,\hspace{0.25in}+\Big\{g^{(2)\mu\nu}\Theta[-\Phi_1]\Theta[\Phi_2]+g^{(3)\mu\nu}\Theta[-\Phi_2]\Big\}\Big\{\Big(\Gamma^{(2)\rho}_{\nu\mu}-\Gamma^{(3)\rho}_{\nu\mu}\Big)\partial_{\rho}\Phi_{2}-\Big(\Gamma^{(2)\rho}_{\rho\mu}-\Gamma^{(3)\rho}_{\rho\mu}\Big)\partial_{\nu}\Phi_{2}\Big\}\delta[\Phi_2].
\end{align}
\ewt

We see that the Ricci tensor and Ricci scalar separate into the Ricci tensor and scalar associated with the three regions as well as terms containing delta function singularities occurring at the stitching surfaces \footnote{The delta function terms in (\ref{riccitcal}) and (\ref{ricciscal}) should be handled with care if we are to treat this expression as a distribution. In general the product $\Theta[x]\delta[x]$ does not make distributional sense as $\theta[x]$ is not continuous at $x=0$. However $\Theta[x]+\theta[-x]=1$ pointwise and hence $\big(\Theta[x]+\Theta[-x]\big)\delta[x]=\delta[x]$ does make distributional sense. Finally since in the global coordinate chart we have $\big(g^{(1)\mu\nu}-g^{(2)\mu\nu}\big)|_{\Phi_1}=0$ and $\big(g^{(2)\mu\nu}-g^{(3)\mu\nu}\big)|_{\Phi_2}=0$, the term $\Big\{g^{(2)\mu\nu}\Theta[-\Phi_1]\Theta[\Phi_2]+g^{(3)\mu\nu}\Theta[-\Phi_2]\Big\}\delta[\Phi_1]$ and the analogous term at $\Phi_{2}=0$ make distributional sense since the coefficient of the delta function is pointwise continuous.}. One thing to note is that care should be taken with two boundary terms. 
For simplicity we  relabel the surface terms

\bwt
\begin{align}
\Delta^{(1)}_{\mu\nu}\delta[\Phi_1] &= \Big\{\Big(\Gamma^{(1)\rho}_{\nu\mu}-\Gamma^{(2)\rho}_{\nu\mu}\Big)\partial_{\rho}\Phi_{1}-\Big(\Gamma^{(1)\rho}_{\rho\mu}-\Gamma^{(2)\rho}_{\rho\mu}\Big)\partial_{\nu}\Phi_{1}\Big\}\delta[\Phi_1]\nonumber\\
\Delta^{(2)}_{\mu\nu}\delta[\Phi_2] &= \Big\{\Big(\Gamma^{(2)\rho}_{\nu\mu}-\Gamma^{(3)\rho}_{\nu\mu}\Big)\partial_{\rho}\Phi_{2}-\Big(\Gamma^{(2)\rho}_{\rho\mu}-\Gamma^{(3)\rho}_{\rho\mu}\Big)\partial_{\nu}\Phi_{2}\Big\}\delta[\Phi_2]\nonumber\\
\Delta^{(1)}\delta[\Phi_1] &= \Big\{g^{(1)\mu\nu}\Theta[\Phi_1]+g^{(2)\mu\nu}\Theta[-\Phi_1]\Theta[\Phi_2]\Big\}\Big\{\Big(\Gamma^{(1)\rho}_{\nu\mu}-\Gamma^{(2)\rho}_{\nu\mu}\Big)\partial_{\rho}\Phi_{1}-\Big(\Gamma^{(1)\rho}_{\rho\mu}-\Gamma^{(2)\rho}_{\rho\mu}\Big)\partial_{\nu}\Phi_{1}\Big\}\delta[\Phi_1]\nonumber\\
\Delta^{(2)}\delta[\Phi_2] &= \Big\{g^{(2)\mu\nu}\Theta[-\Phi_1]\Theta[\Phi_2]+g^{(3)\mu\nu}\Theta[-\Phi_2]\Big\}\Big\{\Big(\Gamma^{(2)\rho}_{\nu\mu}-\Gamma^{(3)\rho}_{\nu\mu}\Big)\partial_{\rho}\Phi_{2}-\Big(\Gamma^{(2)\rho}_{\rho\mu}-\Gamma^{(3)\rho}_{\rho\mu}\Big)\partial_{\nu}\Phi_{2}\Big\}\delta[\Phi_2].
\end{align}

\begin{center}
The Einstein tensor $G_{\mu\nu} =R_{\mu\nu}-\frac{1}{2}R\,g_{\mu\nu}$ can be written as
\end{center}

\begin{align}
G_{\mu\nu}& = R^{(1)}_{\mu\nu}\Theta[\Phi_1]+R^{(2)}_{\mu\nu}\Theta[-\Phi_1]\Theta[\Phi_2]+R^{(3)}_{\mu\nu}\Theta[-\Phi_2] +\Delta^{(1)}_{\mu\nu}\delta[\Phi_1] +\Delta^{(2)}_{\mu\nu}\delta[\Phi_2]-\frac{1}{2}\Big\{R^{(1)}\Theta[\Phi_1]+R^{(2)}\Theta[-\Phi_1]\Theta[\Phi_2]\nonumber\\&\hspace{0.25in}+R^{(3)}\Theta[-\Phi_2]+\Delta^{(1)}\delta[\Phi_1]+\Delta^{(2)}\delta[\Phi_2]\Big\}\Big(g^{(1)}_{\mu\nu}\Theta[\Phi_1]+g^{2)}_{\mu\nu}\Theta[-\Phi_1]\Theta[\Phi_2]+g^{(3)}_{\mu\nu}\Theta[-\Phi_2]\Big)\nonumber\\&=G^{(1)}_{\mu\nu}\Theta[\Phi_1]+G^{(2)}_{\mu\nu}\Theta[-\Phi_1]\Theta[\Phi_2]+G^{(3)}_{\mu\nu}\Theta[-\Phi_2]+\Delta^{(1)}_{\mu\nu}\delta[\Phi_1] +\Delta^{(2)}_{\mu\nu}\delta[\Phi_2]+\big\{g^{(1)}_{\mu\nu}\Theta[\Phi_1]+g^{(2)}_{\mu\nu}\Theta[-\Phi_1]\big\}\Delta^{(1)}\delta[\Phi_1]\nonumber\\&\hspace{.25in}+\big\{g^{(2)}_{\mu\nu}\Theta[\Phi_2]+g^{(3)}_{\mu\nu}\Theta[-\Phi_2]\big\}\Delta^{(2)}\delta[\Phi_2].
\end{align}
\ewt

We see that because of this the Einstein tensor and Ricci scalar break up into their respective values for their regions of spacetime,  i.e.,$G^{(1)}_{\mu\nu}=G^{(3)}_{\mu\nu}=0$, and $G^{(2)}_{\mu\nu}=\Lambda g^{(2)}_{\mu\nu}$; similarly $R^{(1)}=R^{(3)}=0$ and $R^{(2)}=4\Lambda$.

This yields the field equations
\bwt
\begin{align}\label{einsteineom}
G_{\mu\nu}&=G^{(2)}_{\mu\nu}\Theta[-\Phi_1]\Theta[\Phi_2]+\frac{1}{2}\Lambda g^{(2)}_{\mu\nu}\Theta[-\Phi_1]\Theta[\Phi_2]+ \{\Delta^{(1)}_{\mu\nu}+\Delta^{(1)}\big(g^{(1)}_{\mu\nu}\Theta[\Phi_1]+g^{(2)}_{\mu\nu}\Theta[-\Phi_1]\big)\}\delta[\Phi_1]\nonumber\\&\hspace{1.35in} + \{\Delta^{(1)}_{\mu\nu}+\Delta^{(1)}\big(g^{(1)}_{\mu\nu}\Theta[\Phi_2]+g^{(2)}_{\mu\nu}\Theta[-\Phi_2]\big)\}\delta[\Phi_2].
\end{align}
\ewt

The Einstein-Hilbert action that produces this E.O.M. is

\bwt
\begin{align}\label{actioncheck}
S=\frac{1}{2\kappa}\int\,d^{4}x\sqrt{g^{(2)}}\Big(R^{(2)}-2\Lambda\Big)\Theta[-\Phi_1]\Theta[\Phi_2] +S_{\Delta}+S_{boundary},
\end{align}
\ewt

as was argued in the main text. Here, the stress-tensor contributions of the domain wall [the second and third lines of (\ref{einsteineom})] come from the $S_{\Delta}$, which we argued is independent of $\eta_0$ even though the expression of $T_{\mu\nu}$ might have $\eta_0$ dependence depending on the coordinate system.

\section{geodesics and Christoffells}\label{geodchris}
\hspace{0.5in}For reference the geodesic equations of dS in conformal coordinates are
\begin{align}\label{geodesiceqnsds}
\frac{\partial^{2}\eta}{\partial\sigma^{2}}&+\tan{\eta}\Bigg(\frac{\partial\eta}{\partial\sigma}\Bigg)^{2}+\tan{\eta}\Bigg\{\Bigg(\frac{\partial\psi}{\partial\sigma}\Bigg)^{2}+\sin^{2}{\psi}\Bigg[\Bigg(\frac{\partial\theta}{\partial\sigma}\Bigg)^{2}\nonumber\\&\hspace{1.15in}+\sin^{2}{\theta}\Bigg(\frac{\partial\phi}{\partial\sigma}\Bigg)^{2}\Bigg]\Bigg\}=0\\\frac{\partial^{2}\psi}{\partial\sigma^{2}}&+2\tan{\eta}\frac{\partial\eta}{\partial\sigma}\frac{\partial\psi}{\partial\sigma}-\sin{\psi}\cos{\psi}\Bigg[\Bigg(\frac{\partial\theta}{\partial\sigma}\Bigg)^{2}\nonumber\\&\hspace{1.15in}+\sin^{2}{\theta}\Bigg(\frac{\partial\phi}{\partial\sigma}\Bigg)^{2}\Bigg]=0\\\frac{\partial^{2}\theta}{\partial\sigma^2}&+2\tan{\eta}\frac{\partial\eta}{\partial\sigma}\frac{\partial\theta}{\partial\sigma}+2\cot{\psi}\frac{\partial\psi}{\partial\sigma}\frac{\partial\theta}{\partial\sigma}\nonumber\\&\hspace{1.15in}-\sin{\theta}\cos{\theta}\Bigg(\frac{\partial\phi}{\partial\sigma}\Bigg)^{2}=0\\\frac{\partial^{2}\phi}{\partial\sigma^2}&+2\tan{\eta}\frac{\partial\eta}{\partial\sigma}\frac{\partial\phi}{\partial\sigma}+2\cot{\psi}\frac{\partial\psi}{\partial\sigma}\frac{\partial\phi}{\partial\sigma}\nonumber\\&\hspace{1.15in}+2\cot{\theta}\frac{\partial\theta}{\partial\sigma}\frac{\partial\phi}{\partial\sigma}=0.
\end{align}

The geodesic equations for the hats (Milne universe) in the coordinates used in (\ref{hat}) are 
\begin{align}\label{geodesicseqhat}
\frac{\partial^{2}\tau}{\partial\sigma^{2}}&=0\\ \frac{\partial^{2}\psi}{\partial\sigma^{2}}&+2\tan{[\psi+\eta_0]}\Bigg(\frac{\partial\psi}{\partial\sigma}\Bigg)^{2}-\frac{\cos{[\psi+\eta_0]}\sin{\psi}}{\cos{\eta_0}}\nonumber\\&\hspace{0.4in}\times\Bigg[\Bigg(\frac{\partial\theta}{\partial\sigma}\Bigg)^{2}+\sin^{2}{\theta}\Bigg(\frac{\partial\phi}{\partial\sigma}\Bigg)^{2}\Bigg]=0\\\frac{\partial^{2}\theta}{\partial\sigma^2}&+\frac{2\cos{\eta_0}}{\sin{\psi}\cos{[\psi+\eta_0]}}\frac{\partial\psi}{\partial\sigma}\frac{\partial\theta}{\partial\sigma}-\sin{\theta}\cos{\theta}\Bigg(\frac{\partial\phi}{\partial\sigma}\Bigg)^{2}=0\\\frac{\partial^{2}\phi}{\partial\sigma^{2}}&+\frac{2\cos{\eta_0}}{\sin{\psi}\cos{[\psi+\eta_0]}}\frac{\partial\psi}{\partial\sigma}\frac{\partial\phi}{\partial\sigma}+2\cot{\theta}\frac{\partial\theta}{\partial\sigma}\frac{\partial\phi}{\partial\sigma}=0.
\end{align}
\\
\bibliographystyle{apsrev4-1}
\bibliography{bibliography}
\end{document}